%% file: publication.tex
\documentclass[aps,prX,reprint,showpaca]{revtex4-1}

\usepackage[utf8x]{inputenc}

\usepackage[english]{babel}

\usepackage{blindtext}

\usepackage{amsmath,amsthm,amssymb}
\usepackage{bm}

\usepackage{graphicx}
\usepackage[caption=false]{subfig}

\usepackage{hyperref}

\begin{document}

%%%%%%%%%%%%%%%%%%%%%%%%%%%%%%%%%%%%%%%%%%%%%%%%%%%%%%%%%%%%%%%%%%%%%%%%%%%%%%%%
% Frontmatter
%%%%%%%%%%%%%%%%%%%%%%%%%%%%%%%%%%%%%%%%%%%%%%%%%%%%%%%%%%%%%%%%%%%%%%%%%%%%%%%%

\input{front.tex}

%%%%%%%%%%%%%%%%%%%%%%%%%%%%%%%%%%%%%%%%%%%%%%%%%%%%%%%%%%%%%%%%%%%%%%%%%%%%%%%%
% Main body
%%%%%%%%%%%%%%%%%%%%%%%%%%%%%%%%%%%%%%%%%%%%%%%%%%%%%%%%%%%%%%%%%%%%%%%%%%%%%%%%

\input{intro.tex}

\input{theory.tex}

\input{results.tex}

\input{discussion.tex}

%%%%%%%%%%%%%%%%%%%%%%%%%%%%%%%%%%%%%%%%%%%%%%%%%%%%%%%%%%%%%%%%%%%%%%%%%%%%%%%%
% Acknowledgements
%%%%%%%%%%%%%%%%%%%%%%%%%%%%%%%%%%%%%%%%%%%%%%%%%%%%%%%%%%%%%%%%%%%%%%%%%%%%%%%%

\input{acknowledge.tex}

%%%%%%%%%%%%%%%%%%%%%%%%%%%%%%%%%%%%%%%%%%%%%%%%%%%%%%%%%%%%%%%%%%%%%%%%%%%%%%%%
% Appendix
%%%%%%%%%%%%%%%%%%%%%%%%%%%%%%%%%%%%%%%%%%%%%%%%%%%%%%%%%%%%%%%%%%%%%%%%%%%%%%%%

\input{appendix.tex}

%%%%%%%%%%%%%%%%%%%%%%%%%%%%%%%%%%%%%%%%%%%%%%%%%%%%%%%%%%%%%%%%%%%%%%%%%%%%%%%%
% Bibliography
%%%%%%%%%%%%%%%%%%%%%%%%%%%%%%%%%%%%%%%%%%%%%%%%%%%%%%%%%%%%%%%%%%%%%%%%%%%%%%%%

%

\end{document}

%% file: front.tex
\title{Efficient estimation of rare-event kinetics}

\author{Benjamin Trendelkamp-Schroer}
\email{benjamin.trendelkamp-schroer@fu-berlin.de}
\author{Frank No\'{e}}
\email{frank.noe@fu-berlin.de}
\thanks{``corresponding author''}

\affiliation{Institut f\"{u}r Mathematik und Informatik, FU Berlin, Arnimallee 6, 14195 Berlin}

\date{\today}

\begin{abstract}
  The efficient calculation of rare-event kinetics in complex
  dynamical systems, such as the rate and pathways of ligand
  dissociation from a protein, is a generally unsolved problem. Markov
  state models can systematically integrate ensembles of short
  simulations and thus effectively parallelize the computational
  effort, but the rare events of interest still need to be
  spontaneously sampled in the data. Enhanced sampling approaches,
  such as parallel tempering or umbrella sampling, can accelerate the
  computation of equilibrium expectations massively - but sacrifice
  the ability to compute dynamical expectations. In this work we
  establish a principle to combine knowledge of the equilibrium
  distribution with kinetics from fast ``downhill'' relaxation
  trajectories using reversible Markov models. This approach is
  general as it does not invoke any specific dynamical model, and can
  provide accurate estimates of the rare event kinetics. Large gains
  in sampling efficiency can be achieved whenever one direction of the
  process occurs more rapid than its reverse, making the approach
  especially attractive for downhill processes such as folding and
  binding in biomolecules.
\end{abstract}

% \pacs{}

\maketitle

%% file: intro.tex
\section{Introduction}
A wide range of biological or physico-chemical systems exhibit
rare-event kinetics, consisting of rare-transitions between a couple
of long-lived (meta-stable) states. Examples are protein-folding,
protein ligand association, and nucleation processes. Meta-stability
can be found in any system in which states of minimum energy are
separated by barriers higher than the average thermal energy.

A thorough understanding of such systems encompasses the kinetics of
the rare-events, e.g. rates and transition pathways. Obtaining
reliable estimates for such systems is notoriously difficult: The
simulation time needs to exceed the longest waiting time, resulting in
a sampling problem.

In recent years, Markov state models (MSMs) \cite{huisinga1999,
  swope2004, singhal2004, ChoderaEtAl_JCP07,
  NoeHorenkeSchutteSmith_JCP07_Metastability, pan2008,
  BucheteHummer_JPCB08, prinz2011} and their practical applicability
through software \cite{senne2012, BeauchampEtAl_MSMbuilder2} have
become a key technology for computing kinetics of complex rare-event
systems. A well-constructed MSM separates the kinetically distinct
states and captures their transition rates or probabilities. With a
suitable choice of state space discretization and lag-time, kinetics
can be approximated with high numerical accuracy
\cite{SarichNoeSchuette2010, prinz2011}.
MSMs can be straightforwardly interpreted and analyzed with Markov
chain theory and Transition Path Theory \cite{vanden2006, metzner2009}.
This was demonstrated, for example, for protein folding \cite{noe2009,
  VoelzPande_JACS10_NTL9} or protein-ligand binding \cite{held2011,
  huang2014}. 

MSMs can somewhat alleviate the sampling problem by
virtue of the fact that they can be estimated from short simulations
produced in parallel \cite{noe2009, VoelzPande_JACS10_NTL9,
  BuchFabritiis_PNAS11_Binding}, thus avoiding the need for single
long trajectories \cite{shaw2010}. However, the rare-events of
interest must be sampled in the data in order to be captured by the
model. For example, in protein-ligand binding, a dissociation rate can
only be computed if each step of the dissociation process has been
sampled at least once.

Orders of magnitude of speedup can be achieved with enhanced sampling
methods such as umbrella sampling, replica exchange dynamics, or
meta-dynamics \cite{torrie1977, grubmueller1995, sugita1999,
  laio2002}. The speedup is achieved by coupling the unbiased ensemble
of interest with ensembles at higher temperature at which the
rare-events occur more frequently or by using biasing potentials
allowing to ``drag'' the system across an energy barrier. With such
approaches, accurate equilibrium expectations, such as free energy
profiles can be computed efficiently, but the dynamical properties of
the unbiased ensemble, such as transition rates, relaxation
time-scales and transition pathways, are generally not available.

A common approach to reconstruct the kinetics from the free energy
profiles is to employ rate theories such as transition state theory,
Kramer's or Smoluchowski/Langevin models \cite{eyring1935,
  kramers1940, BestHummer_PNAS09_Diffusion, tiwary2013}. Such
dynamical models introduce additional assumptions that cannot be
self-consistently validated because the predicted dynamics is not
present in the data. 

A much more advanced approach was recently introduced in \cite{rosta2015},
where an MSM-based estimator for the stationary vector using transition counts
harvested from simulations at different thermodynamic states, such as umbrella
sampling. This approach allows to mix, in principle, umbrella sampling
simulations and direct molecular dynamics and compute the rates from the
transition matrix of the unbiased ensemble. However, the coefficient matrices used 
to connect the biased and unbiased transition matrices require a specific 
dynamical model to be formulated (such as Brownian dynamics in the
free energy coordinate), making this approach essentially a rate model.
In general, key assumptions underlying rate models are usually the existence 
of a time-scale separation and approximate Markovianity on a single or 
few reaction coordinates - assumptions that are unlikely to hold for 
complex multi-state systems describing, e.g. bio-molecular dynamics.

The recently introduce transition-based reweighting analysis methods
(TRAM) \cite{WuNoe2014, WuMeyRostaNoe2014, MeyWuNoe2014} permit to
rigorously combine direct molecular dynamics and enhanced sampling
methods towards full thermodynamics and kinetics without assuming any
restrictive rate model. However, a current limitation is that in order
to extract unbiased kinetics, the transition events need to be
evaluated at a common and sufficiently large lag time $\tau$ at all
thermodynamic states. This requirement is not consistent with efficient
umbrella sampling or replica-exchange MD simulations that typically
employ very short simulation snippets.

Finally, computation of kinetic quantities without rate models is also possible 
with path sampling methods, such as transition path sampling
\cite{bolhuis2002}, milestoning \cite{faradjian2004}, transition
interface sampling \cite{vanErp2003}, and multi-state transition
interface sampling \cite{du2013}. A challenge is that these approaches are
essentially two-state methods. The transition end-states must be defined a-priori and
all relevant rare-events must be distinguishable in the reaction
coordinates, cores, or milestones that the method operates on.

Here we construct a general simulation approach that enables the
computation of kinetic observables related to slow processes without
having to explicitly sample the rare-events. It is based upon a very
simple but general idea: Simulations are often constructed in such a
way that they obey microscopic reversibility or at least a
generalization thereof \cite{vanGunsteren1988, tuckerman1992}. In this
case, for any partition of state space into sets $i,j,\dotsc$ and any
choice of the lag-time $\tau$, we have the detailed balance relation
\begin{equation}
  \label{eq:detailed_balance_intro}
  \pi_i p_{ij}(\tau) = \pi_{j} p_{ji}(\tau),
\end{equation}
where $p_{ij}(\tau)$ is the probability of making a transition from
set $i$ to set $j$ within a time $\tau$ and $\pi_i$ is the equilibrium
probability of set $i$. Suppose we have knowledge about the
equilibrium probabilities $\pi_i$, $\pi_j$ from an enhanced sampling
simulation. Then only the larger one of the two transition
probabilities - $p_{ij}$ or $p_{ji}$ - needs to be sampled while the
less probable event can be reconstructed by
\eqref{eq:detailed_balance_intro}. 

Speaking in terms of a network of states, a direct analysis or an
analysis via Markov state models requires all states to be connected
in both directions (strongly connected). The presented method allows
to relax this requirement if an estimate of the equilibrium
probabilities is given - now all states need to be only connected in
one direction (weakly connected).

The slow rate exhibits a functional dependence on the transition
probabilities of the slow event. By virtue of the detailed balance
condition a reliable estimate of the transition probabilities for the
frequent event entails a reliable estimate for the transition
probabilities of the slow event - resulting in a reliable estimate of
the slow rate.

While the inference procedure is trivial for a two-state system where
three of the four components in \eqref{eq:detailed_balance_intro} are
known exactly, it is far from trivial for a system with many states
and when some or all estimates are subject to statistical
uncertainty. Here we establish a systematic inference scheme for
combining multi-state estimates of the equilibrium probabilities
$(\pi_i)$ with sampling data of at least the ``down-hill'' transition
probabilities $p_{ij}$. Our approach is built upon the framework of
reversible Markov models \cite{SriramanKevrekidisHummer_JPCB109_6479,
  noe2008} where \eqref{eq:detailed_balance_intro} is enforced between
all pairs of states. As a consequence, our estimates do not invoke any
additional dynamical model, are accurate within a suitable state space
discretization \cite{SarichNoeSchuette2010, prinz2011}, and are
precise in the limit of sufficient sampling.

In contrast to the DHAM and TRAM methods we use the stationary
probabilities already estimated from enhanced sampling simulation as
additional input parameters for the estimation of MSM transition
probabilities. Standard reweighting schemes used to obtain the
stationary probabilities do usually not assume a dynamical model to
obtain the desired unbiased probabilities.

The estimation procedure can reduce the sampling problem tremendously
for processes with some long-lived states and some other states from
which the system relaxes rapidly. This case is ubiquitous in
meta-stable systems, because long-lived states are connected by
short-lived transition states. But even long-lived states usually have
very different lifetimes: For example many ligands or inhibitors bind
to their protein receptor with nanomolar concentrations, meaning
that the transition probabilities leading to the associated state are
orders of magnitude higher than the dissociation probabilities. The
present reversible Markov model approach lays the basis for estimating
the kinetics and mechanisms of protein-drug dissociation by combining
the much more rapid association trajectories with suitable enhanced
sampling methods such as Hamiltonian replica-exchange \cite{wang2013}
or umbrella sampling \cite{SouailleRoux_CPC01_WHAM,
  BuchFabritiis_PNAS11_Binding}.

%% file: theory.tex
\section{Theory}

\subsection{Markov state models}
Classical dynamics, governed by Newton's equations in the case of an
isolated system and by Langevin equation's for systems at constant
temperature \cite{zwanzig1973}, gives rise to a transfer operator
$\mathcal{P}$ propagating a phase-space density from time $t$ to time
$t+\Delta t$ \cite{gardiner1986, schuette1999,
  schuette2000}. Numerical solutions for Newton's or Langevin
equations can be obtained for complex systems with many degrees of
freedom, but a direct numerical assessment of the transfer operator is
in most cases prohibited due to the curse of dimensionality.

Markov state models (MSMs) bridge this gap estimating the transfer
operator on a suitably defined state space partition
\begin{equation}
  \label{eq:state_space}
  \Omega=\{s_1,\dotsc,s_n\}.
\end{equation}
using trajectories obtained by direct numerical simulation
\cite{SarichNoeSchuette2010, prinz2011}.

MSMs model the jump process between states of this partition by a
Markov chain. Observed transitions between pairs of states $i$ and $j$
are collected in a \emph{count matrix} $C=(c_{ij})$ and the likelihood
for the observed counts for a given \emph{transition matrix}
$P=(p_{ij})$ is given by
\begin{equation}
  \label{eq:likelihood_multinomial}
  \mathbb{P}(C|P) \propto 
  \prod_{i}\left(\prod_{j} p_{ij}^{c_{ij}} \right).
\end{equation}

While the likelihood functions allows to determine the maximum
likelihood estimator $\hat{P}$ optimizing the likelihood function for
a given observation $C$ over the set of all possible models $P$ it
does not specify the uncertainty of a chosen model.

For a finite amount of observation data there will in general be a
whole ensemble of models compatible with the given data. In order to
specify uncertainties and determine statistical errors of estimated
quantities we need to infer the posterior probability of a model for a
given observation. An application of Bayes' formula yields
\begin{equation}
  \label{eq:Bayes_formula}
  \underbrace{\mathbb{P}(P|C)}_{posterior} \propto
  \underbrace{\mathbb{P}(P)}_{prior}
  \underbrace{\mathbb{P}(C|P)}_{likelihood}.
\end{equation}
For a uniform prior, i.e. no a priori knowledge about the model, the
posterior probability is given as a product of \emph{Dirichlet
  distributions}
\begin{equation}
  \label{eq:Dirichlet_posterior}
  \mathbb{P}(P|C) \propto \prod_{i} \left( \prod_j p_{ij}^{c_{ij}} \right).
\end{equation}

\subsection{Inference using a given stationary vector}
There are many methods that allow to efficiently estimate the
stationary vector, even in situations in which a direct estimation
from a finite observation of the Markov chain is unfeasible due to the
meta-stable nature of the system \cite{torrie1977, grubmueller1995,
  sugita1999, wang2001, laio2002, trebst2006}. In such situations it
is often possible to alter the system dynamics in a controlled way
such that the artificial dynamics equilibrates more rapidly than the
original one. The desired stationary vector of the original dynamics
can then be related to the stationary vector estimated from the
altered process \cite{bennett1976, ferrenberg1989, kumar1995, tan2004, shirts2008}.

In the following we want to show how such prior knowledge about the
stationary vector can be used to improve the estimates of kinetic
observables in systems with rare-events.

We are again given a finite observation of a Markov chain in terms of
the count matrix $C$. Assume we are additionally given the stationary vector
$\pi$ for our system of interest and we know that the transition
probabilities of the chain fulfil detailed balance for the given
stationary vector,
\begin{equation}
  \label{eq:defn_detailed_balance}
  \pi_i p_{ij} = \pi_j p_{ji}.
\end{equation}
Then we can express the posterior probability for our model via
\eqref{eq:Bayes_formula}. Prior knowledge about the stationary vector
$\pi$ in combination with the detailed balance assumption formally entails the
following prior distribution on the posterior ensemble,
\begin{equation}
  \label{eq:prior_fixed_pi}
  \mathbb{P}(P|\pi)=\prod_{i<j} \delta\left(\pi_i p_{ij} - \pi_j p_{ji} \right).
\end{equation}
According to \eqref{eq:Bayes_formula} the constrained posterior is
\begin{equation}
  \label{eq:posterior_fixed_pi}
  \mathbb{P}(P|C,\pi) \propto \mathbb{P}(P|\pi) \mathbb{P}(C|P).
\end{equation}
The effect of the prior \eqref{eq:prior_fixed_pi} is a restriction of
the posterior to the subspace of transition matrices fulfilling
detailed balance with respect to the fixed stationary vector $\pi$.

\subsection{Maximum likelihood estimate given a stationary vector}
We can also use prior knowledge of the stationary vector to constrain
the maximum likelihood estimate $\hat{P}$ to the set of matrices
obeying \eqref{eq:defn_detailed_balance} for a given stationary vector
$\pi$. This results in the following convex constrained optimization
problem
\begin{equation}
  \label{eq:mle_constrained}
  \begin{aligned} &\text{minimize} & -\sum_{i,j} c_{ij} \log p_{ij} \\
    &\text{subject to} & p_{ij} \geq 0 \\
    &                  & \sum_{j} p_{ij}=1 \\
    &                  & \pi_{i}p_{ij}=\pi_{j}p_{ji}
    \end{aligned}
\end{equation}
which can be solved using methods outlined in
\cite{TrendelkampWuNoe2015}.

\subsection{Inference using a stationary vector with uncertainty}
A stationary vector estimate usually carries a finite sampling error
which should be accounted for when inferring a reversible transition
matrix from data. From a Bayesian viewpoint we have to combine two
sources of evidence. The observed count-matrix $C$ from standard
equilibrium simulations and the data from enhanced or biased sampling
methods $E$ used to estimate the stationary vector.

An error model for the estimation of uncertainty in the stationary
vector assess the posterior of stationary vectors given the enhanced sampling data, $\mathbb{P}(\pi|E)$.  Recent methods for the
uncertainty quantification of reversible MSMs with fixed stationary
vector allow to sample the posterior $\mathbb{P}(P|\pi, C)$ in
\eqref{eq:posterior_fixed_pi}. 

The posterior for transition matrices
under the combined evidence $\mathbb{P}(P|C,E)$ can be formally
decomposed as
\begin{equation}
  \label{eq:posterior_combined_evidence}
  \mathbb{P}(P|C, E)=\int \mathrm{d} \pi \, \mathbb{P}(P|C,\pi,E)\mathbb{P}(\pi|C,E).
\end{equation}
Assuming that the direct effect of the enhanced sampling
information $E$ is negligible in the posterior of transition matrices
with given stationary vector,
\begin{equation}
  \label{eq:approx_1}
  \mathbb{P}(P | C,\pi,E) \approx \mathbb{P}(P | C, \pi)
\end{equation}
and that the direct effect of observed transition counts $C$ is
unimportant compared to the enhanced sampling data used to obtain
$\pi$ from a standard reweighting scheme,
\begin{equation}
  \label{eq:approx_2}
  \mathbb{P}(\pi | C, E) \approx \mathbb{P}(\pi | E)
\end{equation}
we model the uncertainty encoded in the desired posterior by inserting
the two approximations \eqref{eq:approx_1}, \eqref{eq:approx_2} into
\eqref{eq:posterior_combined_evidence}
\begin{equation}
  \label{eq:posterior_combined_evidence_approx}
  \mathbb{P}(P | C, E) \approx \int \mathrm{d} \pi \, \mathbb{P}(P | C, \pi) \mathbb{P}(\pi | E).
\end{equation}

Approximate sampling from $\mathbb{P}(P|C, E)$ can now be achieved by
drawing a random sample $\pi^{(1)}, \dotsc, \pi^{(M)}$ distributed
according to a given error-model, $\pi^{(k)} \sim \mathbb{P}(\pi|E)$
and generating a sample of transition matrices
$P_1^{(k)},\dotsc,P_N^{(k)}$ from the constrained posterior $P_i^{(k)}
\sim \mathbb{P}(P|C,\pi^{(k)})$ for each of the $\pi^{(k)}$. The
sample
$P_1^{(1)},\dotsc,P_{N}^{(1)},\dotsc,P_1^{(M)},\dotsc,P_N^{(M)}$ will
then be approximately distributed according to $\mathbb{P}(P|C, E)$.

In \cite{noe2008} we have presented a Markov chain Monte Carlo
approach to sample reversible transition matrices fulfilling detailed balance
with respect to a fixed stationary vector. This method however has
suffered from poor acceptance probabilities. In
\cite{TrendelkampWuNoe2015} we outline a method to efficiently generate
samples from the constrained posterior using a Gibbs sampling
algorithm that will be used here.

For given vector $(\pi_i)$ detailed balance
\eqref{eq:defn_detailed_balance} enforces a linear dependence between
the transition matrix element $p_{ij}$ and the element $p_{ji}$. As an
immediate consequence the standard error of both elements for a sample
generated from the posterior $\mathbb{P}(P|C)$ has to be equal,
\begin{equation}
  \label{eq:fundamental_error_relation}
  \frac{\sqrt{\mathbb{V}(p_{ji})}}{\mathbb{E}(p_{ji})} =   \frac{\sqrt{\mathbb{V}(p_{ij})}}{\mathbb{E}(p_{ij})}.
\end{equation}
We will show how this can be used in order to significantly improve
various estimates in situations in which $p_{ij} \ll p_{ji}$.

%% file: results.tex
\section{Results}
In the following we will demonstrate the usefulness of
\eqref{eq:fundamental_error_relation} via a comparison of the standard
error for kinetic quantities depending on rare-events that are either
estimated from a Markov model of the direct unbiased simulation
(unconstrained posterior \eqref{eq:Dirichlet_posterior}), as well as
when enhanced sampling data is additionally used (constrained
posterior \eqref{eq:posterior_fixed_pi} or constrained posterior with
uncertain stationary vector
\eqref{eq:posterior_combined_evidence_approx}).

\subsection{Finite state space Markov chain}
Consider a three-state Markov chain with the following transition matrix
\begin{equation}
  \label{eq:model_system}
  P=\left(
    \begin{array}{ccc}
      1-10^{-b} & 10^{-b} & 0 \\
      \frac{1}{2} & 0 & \frac{1}{2} \\
      0 & 10^{-b} & 1-10^{-b}
    \end{array}
  \right).
\end{equation}
The parameter $b>0$ can be thought of as the height of an energy
barrier between states one and three. The corresponding stationary
distribution is given by
\begin{equation}
  \label{eq:stationary_distribution_toy_model}
\pi=(1+10^{-b})^{-1} \left(\frac{1}{2}, 10^{-b}, \frac{1}{2}\right)^T.
\end{equation}
The pair $(\pi, P)$ satisfies the detailed balance equation
\eqref{eq:defn_detailed_balance}.

Any process starting in state one has an exponential small probability
of crossing over to state three. In fact a chain starting in state one
can reach state three only via state two, but the probability to go
from state one to state two is exponentially small in the barrier
height $b$. The reversed process, going from state two to state one,
occurs much faster. The same applies to state three and state
two. The eigenvalues of this matrix are
\begin{equation*}
  \begin{aligned}
    & \lambda_1=1 & \lambda_2=1-10^{-b} & & \lambda_3=-10^{-b}. 
  \end{aligned}
\end{equation*}
and the slowest time-scale in the system is given by,
\begin{equation*}
  t_2= -\frac{1}{\log \lambda_2} \approx 10^{b}.
\end{equation*}

It is apparent from $t_2 \approx p_{12}^{-1}$ that estimates of $t_2$
and of $p_{12}$ have similar standard errors. The standard error
$\epsilon$ for a matrix-element $p_{ij}$ for sampling from the
unconstrained posterior \eqref{eq:Dirichlet_posterior} is
\begin{equation*}
  \epsilon(p_{ij})=\frac{1}{\sqrt{c_{ij}}}.
\end{equation*}
For $b=4$ and a single chain of length $N \approx 7 \cdot 10^{4}$
steps starting in state one we can on average expect $c_{12}=4$
resulting in a relative standard error of $50\%$. In order to decrease
the error down to $1\%$ we would need to run a chain of length $N
\approx 100^2 \cdot 10^{4}=10^{8}$ steps. This is clearly an
unsatisfactory situation and we would like to reduce the required
simulation effort to reach a given error level as much as possible.

In comparison for an ensemble of $M$ short chains of length $L$, with
$L \ll 10^{b}$, starting in state two one will on average observe a
transition from state two to state one for every second chain,
$c_{21}=M/2$, so that a relative error of $1\%$ for $p_{21}$ can
already be achieved for $M \approx 10^{4}$, with $L \ll 10^{b}$, so
that the total simulation effort can be reduced by orders of
magnitude.

We do not have explicit expressions for the standard error of matrix
elements $p_{ij}$ when sampling from the restricted ensemble enforcing
detailed balance with respect to a given stationary vector. It is
however conceivable that the standard errors of $p_{21}$ can be
reduced in the same way. The relation
\eqref{eq:fundamental_error_relation} guarantees that a small error
for $p_{21}$ will also result in a small error for the rare-event
quantity $p_{12}$.

\begin{figure}
  \includegraphics[width=\columnwidth]{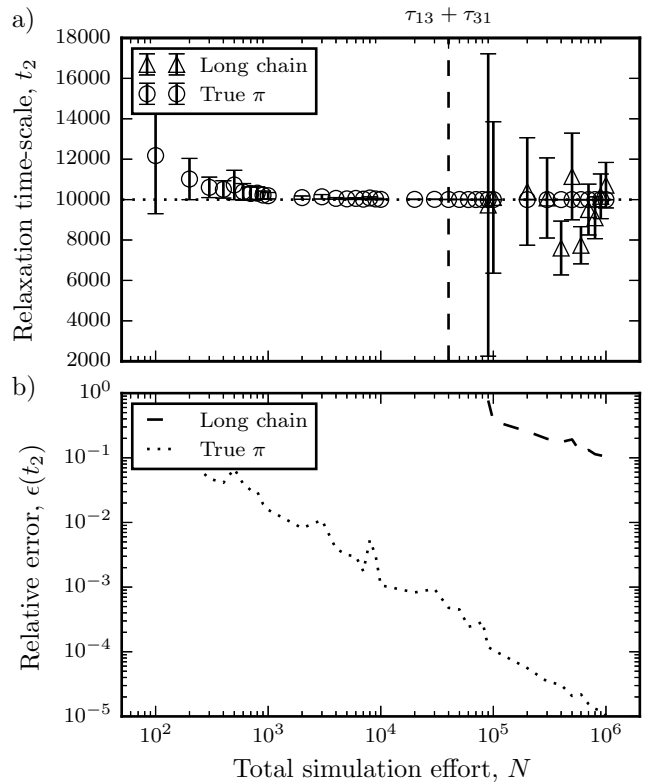}
  \caption{Mean and standard error of the largest implied time-scale
    $t_2$, total simulation effort $N$ for meta-stable 3-state system with barrier parameter $b=4$. a) Convergence of the mean value, using either a single
    long trajectory starting in one of the meta-stable states or the
    stationary vector together with an ensemble of short chains
    relaxing from the transition state. The latter approach allows to
    obtain a reliable estimate already before the average waiting time
    for a single rare-event $\tau_{13}+\tau_{31}$ has elapsed. The
    comparison of the estimated standard error b) indicates a three
    orders of magnitude speedup when estimating the rare-event
    sensitive quantity $t_2$ using the stationary vector in
    combination with short relaxation trajectories.}
  \label{fig:discrete}
\end{figure}

\autoref{fig:discrete} shows the standard error of $t_2$ versus the
total simulation effort. The error for a single long chain is
estimated from a sample of transition matrices generated from the
unconstrained posterior. The error for the ensemble of short chains is
estimated from a sample of transition matrices generated from the
constrained posterior using the algorithm outlined in
\cite{TrendelkampWuNoe2015}. From \autoref{fig:discrete} it is apparent
that using a-priori information about the stationary vector in
combination with an ensemble of short simulations started from the
unstable state results in a three orders of magnitude smaller
simulation effort when trying to estimate $t_2$ with a prescribed
error. In particular, estimation of the rare-event kinetics can be
conducted orders of magnitude before a direct simulation would even
encounter a single transition event.

\begin{figure}
  \centering
  \includegraphics[width=\columnwidth]{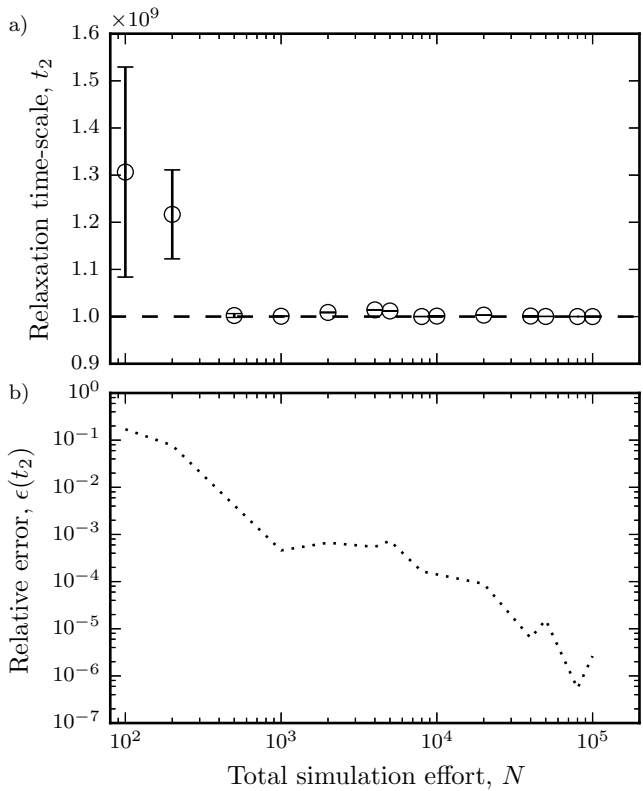}
  \caption{Mean and standard error of the largest relaxation
    time-scale, $t_2$, total simulation effort $N$ for meta-stable
    3-state system with barrier parameter $b=9$. a) Convergence of the
    mean value using short trajectories relaxing from the transition
    state. A correct estimate can be obtained six orders of magnitude
    before a single rare-event would have occurred on average. b)
    Standard error of the estimate. The estimation using long trajectories is unfeasible.
  }
  \label{fig:discrete_b9}
\end{figure}

This effect is even more pronounced when choosing $b=9$ so that
estimation via long trajectories sampling the rare event is
hopeless. Using short trajectories starting in the transition state in
combination with the stationary vector one can accurately estimate
$t_2$ with a total simulation effort of $N=10^{3}$ steps,
c.f. \autoref{fig:discrete_b9}. That is six orders of magnitude before
on average even a single rare-event would have been observed.

\subsection{Double-well potential}
\begin{figure}
  \centering
  \includegraphics[width=\columnwidth]{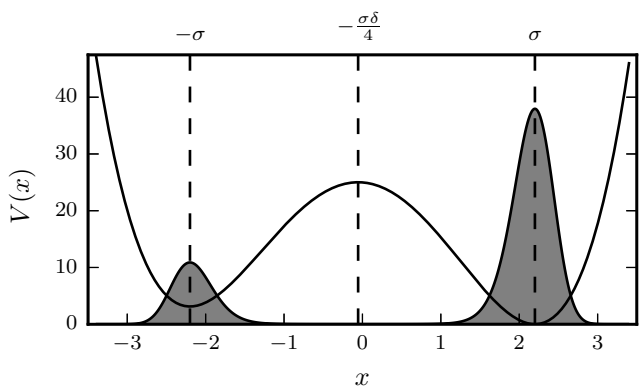}
  \caption{Potential $V(x)$ and stationary distribution $\pi(x)$,for
    Brownian dynamics in double-well potential. The stationary
    distribution (shaded area) is scaled to fit the scale of the
    potential function. It can be seen that the stationary probability
    is concentrated in the meta-stable regions around the two minima
    of the potential at $\pm \sigma$.}
  \label{fig:doublewell_potential}
\end{figure}

Let us now go to an example where the Markov state model is an
approximation of the true dynamics. We employ Brownian dynamics in a
double-well potential defined by
\begin{equation}
  \label{eq:doublewell_potential}
  V(x)=(x^2-\sigma^2)^2+\delta \sigma (\frac{1}{3} x^3 - \sigma^2 x).
\end{equation}
The two minima of the potential at $\pm \sigma$ are separated by a
maximum at $-\delta \sigma/4$,
cf. \autoref{fig:doublewell_potential}. The dynamics is governed by
the following SDE,
\begin{equation}
  \label{eq:Brownian_dynamics}
  \mathrm{d}X_t=-\nabla V(X_t)+ \sqrt{2 \beta^{-1}} \mathrm{d}W_t,
\end{equation}
with $\mathrm{d}W_t$ denoting the increments of the
\emph{Wiener-process}.  The inverse temperature $\beta=(k_B T)^{-1}$
controls the intensity of the stochastic fluctuations.

\eqref{eq:Brownian_dynamics} defines a process where $X_t$ sample from
the canonical distribution,
\begin{equation}
  \label{eq:canonical_ensemble}
  \pi(x)=Z(\beta)^{-1} e^{-\beta V(x)}.
\end{equation}
The temperature dependent constant $Z(\beta)$ is the \emph{partition
function} ensuring correct normalization, $\int \mathrm{d}x \,
\pi(x)=1$. Spectral properties of this Markov process, such as the
largest implied time-scale can be computed from a spatial
discretisation of its associated transition kernel,
cf. \autoref{sec:transition_kernel}. 

For the numerical experiment we used a double-well potential with
parameters $\sigma=2.2$ and $\delta=0.1$. The time step for the
explicit Euler-scheme is $\Delta t=10^{-3}$. The noise parameter is
$\beta=0.4$. 

Spatial discretization of the transition kernel is performed with
$L_x=3.4$ and $n_x=400$ regular sub-intervals. The matrix $(p_{ij})$ is
assembled by evaluating the kernel at the midpoints of the
sub-intervals. The largest implied time scale, $t_2=1.2\cdot 10^{6}$, is
computed from an eigenvalue decomposition of the assembled
matrix. Mean first passage times between sets $A=[\sigma-0.2,
\sigma+0.2]$ and $B=[-\sigma-0.2, -\sigma+0.2]$ are computed as
$\tau_{AB}=5.3 \cdot 10^{6}$ and $\tau_{BA}=1.6 \cdot 10^{6}$, see
\autoref{sec:mfpt} for details. Values computed from the spatial
discretization are used as reference values for comparison with
estimates obtained from a Markov model.

The Markov model is build using a regular grid discretization of $[-L,
L]$ with $L=3.4$ and $n=100$ states. From an implied time scale
estimation using long trajectories with $N=10^8$ steps we obtain
a lag-time of $\tau=10 \, dt$. 

The stationary vector is estimated from umbrella sampling simulations
using the weighted histogram analysis method \cite{ferrenberg1989,
  kumar1992}. Estimates were computed using $M_{\pi}=20$ umbrella sampling
simulations with $L_{\pi}=2.5\cdot 10^{4}$ points per umbrella as well as
from umbrella sampling simulations with $L_{\pi}=5 \cdot 10^{6}$ points per
umbrella. To account for the uncertainty in the estimated stationary
vector we used bootstrap resampling \cite{efron1994} of the generated
data and computed the stationary distribution for each re-sampled data
set to model the ensemble of stationary vectors compatible with the
observed umbrella sampling data.

\begin{figure}
  \includegraphics[width=\columnwidth]{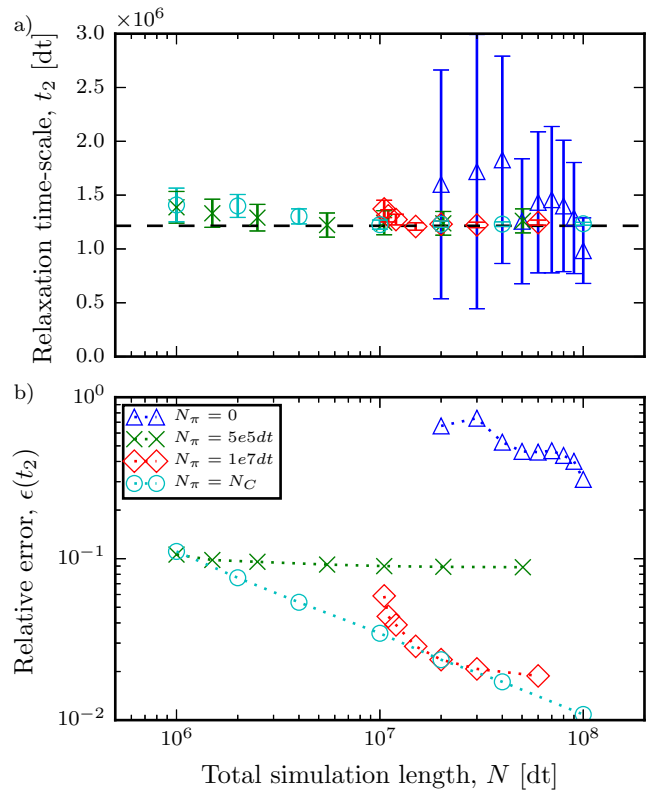}
  \caption{Mean and standard error of largest implied time-scale
    $t_2$, given total simulation effort $N$, for Brownian dynamics in
    double-well potential. a) Convergence of the mean value, using
    either a single long trajectory starting in one of the meta-stable
    states or the stationary vector together with an ensemble of short
    chains relaxing from the transition state. The latter approach
    allows to obtain a reliable estimate already before the average
    waiting time for a single rare-event $\tau_{AB}+\tau_{BA}$ has
    elapsed. A comparison of the standard error b) indicates a more
    than two orders of magnitude speedup when estimating the
    rare-event sensitive quantity $t_2$. By combining short trajectories with
    information about the stationary probabilities, reliable estimates of the
    slowest relaxation timescale can be obtained with a total amount of
    simulation data that is about one order of magnitude smaller than 
    the expected waiting time for a forward and backward transition across the barrier.}
  \label{fig:doublewell_t2}
\end{figure}

In \autoref{fig:doublewell_t2} we show mean and standard error of the
largest implied time scale $t_2$ versus the total simulation effort
$N$. The total simulation effort $N$ is composed of the simulation
effort spent on obtaining a count matrix from standard simulations,
$N_{C}$, and the simulation effort spent on obtaining the stationary
distribution from umbrella sampling simulations, $N_{\pi}$,
\begin{equation}
  \label{eq:total_effort}
  N=N_{\pi}+N_{C}.
\end{equation}

We compare three different approaches when estimating mean and standard
error of the largest implied time-scale $t_2$.
\begin{enumerate}
\item Generate a single trajectory starting in one of the meta-stable
  regions and compute estimates without a priori knowledge of the
  stationary vector.
\item Generate an ensemble of short trajectories starting on the barrier
  and compute estimates with an error-model for the stationary vector
  as prior information.
\item Balanced sampling: split the total simulation effort equally
  between umbrella simulations and short trajectories starting on the
  barrier, $N_{\pi}=N_{C}=N/2$. Compute estimates updating the error
  model for the stationary vector according to the increasing amount
  of data available for the estimation.
\end{enumerate}
Transition matrices are sampled according to
\eqref{eq:Dirichlet_posterior} if no prior knowledge about the
stationary vector is available and from
\eqref{eq:posterior_combined_evidence_approx} if the stationary vector
was estimated from umbrella simulation data. For the first approach we
use $M_{C}=20 \dots 100$ long trajectories of length $L_{C}=10^{6} \, dt$
starting in the minimum point, $x_0=s$, and for the second approach we
use an ensemble of $M_{C}=50 \dots 5000$ short trajectories of length
$L_{C}=10^{4} \, dt$ starting on the barrier, $x_0=-\delta \sigma/4$. For the
second approach we have estimated the stationary vector from a small
as well as for a large amount of umbrella sampling data in order to
demonstrate the dependence of the standard error of the kinetic
observable on the error in the ensemble of input stationary
distributions.

It can be seen from \autoref{fig:doublewell_t2} that for a fixed
effort $N_{\pi}=M_{\pi} L_{\pi}$ the standard error can not be reduced
below a certain amount with increasing $N_{C}=M_{C} L_{C}$. This is a
result of the non-zero statistical error in the estimate of the
stationary vector for fixed $N_{\pi}$. The usual $N^{-\frac{1}{2}}$
dependence of the standard error can be recovered for the proposed
splitting $N_{\pi}=N_{C}=N/2$. \autoref{fig:doublewell_t2} shows the
favourable scaling coefficient of such an approach leading to a more
than two orders of magnitude faster convergence of the estimated quantity
compared to using standard simulations alone. Reliable estimates of
the rare-event kinetics can be obtained one order of magnitude
simulation effort before the standard approach using long trajectories
and no information about the equilibrium probabilities can be applied
at all. The finite error for the estimate of the stationary vector for
$N_{\pi} = 5 \cdot 10^{4} \, dt$ and $N_{\pi} = 10^{7} \, dt$ results in a
saturation of the error of $t_2$ which can be further decreased using
a more precise estimate of the stationary vector from additional
enhanced sampling simulations.

For metastable systems we propose the following strategy for
distributing initial conditions exploiting the information from the
equilibrium vector. Once all meta-stable sets and all kinetic barriers
separating the sets have been identified using some enhanced
sampling protocol, short trajectories should be started on top
of all barriers or in high-energy metastable states. The length of the 
short trajectories needs to be sufficient to relax towards the low-energy
metastable states. The method described here can be used to combine
these data to an estimate of the full rare-event kinetics.

\subsection{Alanine dipeptide}
\begin{figure}
  \includegraphics[width=\columnwidth]{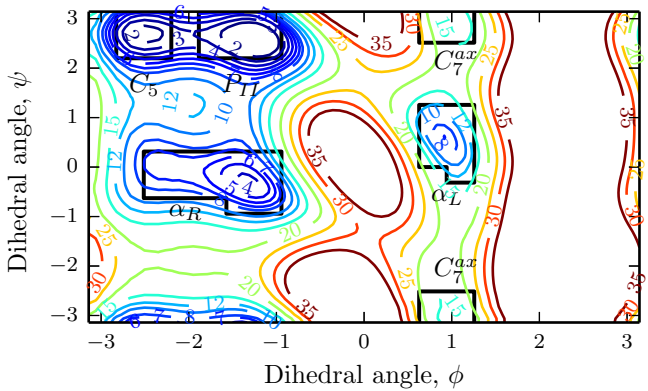}
  \caption{Free energy profile of alanine-dipeptide as a function of
    the dihedral angles. Energies are given in $kJ/mol$. The average
    thermal energy $k_{B}T$ at $300K$ is $2.493 kJ/mol$. One can
    identify five meta-stable sets on the dihedral angle torus, here
    indicated by black lines. There are three low energy (high
    probability) sets $C_5$, $P_{II}$ and $\alpha_R$ with $\phi<0$ and
    two high energy (low probability) sets $\alpha_R$ and $C_7^{ax}$
    with $\phi>0$.}
  \label{fig:pmf_alanine}
\end{figure}

As an example for a rare-event quantity in a molecular system we use the
mean first-passage time for the $C_5$ to $C_7^{ax}$ transition in the
alanine-dipeptide molecule. Alanine-dipeptide has been the
long-serving laboratory rat of molecular dynamics
\cite{montgomery1985, anderson1988, tobias1992, chodera2006,
  du2011}. The $\phi$ and $\psi$ dihedral angles have been identified
as the two relevant coordinates for the slowest kinetic processes of
the system in equilibrium. The potential of mean force for the two
dihedral angles is shown in \autoref{fig:pmf_alanine}.

One can identify five meta-stable regions in the free-energy
landscape. The $C_5$ and $P_{II}$ regions correspond to dihedral
angles found in a beta-sheet conformation, the $\alpha_{R}$ and
$\alpha_{L}$ regions correspond to a right, respectively left-handed
$\alpha$-helix conformation. Reference values for the mean-first
passage times between all pairs of sets have been computed from the
maximum likelihood estimator of \eqref{eq:likelihood_multinomial}
using a total of $10 \mu s$ of simulation data. Values can be found in
\autoref{tab:alanine_mfpt_ref}. For details of the computation of mean
first-passage times see \autoref{sec:mfpt}.

\begin{table}
  \begin{ruledtabular}
  \begin{tabular}{rrrrrr}    
    $\tau_{AB}/ns$      & $C_5$   & $P_{II}$ & $\alpha_R$  & $\alpha_{L}$ & $C_7^{ax}$ \\ 
    $C_5$       &  0     &  0.021 &   0.253  &  43.456 & 60.220 \\
    $P_{II}$     &  0.041 &  0       & 0.255  &  43.449 & 60.213 \\
    $\alpha_R$  & 0.142 &  0.125   & 0      &  43.549 & 60.312 \\
    $\alpha_L$ & 1.553 & 1.527    & 1.744  & 0       & 17.757 \\
    $C_{7}^{ax}$ & 1.559 & 1.533 & 1.745  & 1.221  & 0 \\
  \end{tabular}
  \end{ruledtabular}
  \caption{Mean first passage time (mfpt) between meta-stable regions
of alanine-dipeptide. The mfpts have been estimated from $10 \mu s$ of
simulation data using a Markov state model.}
  \label{tab:alanine_mfpt_ref}
\end{table}

All computations were carried out on high-performance GPU cards using
the OpenMM simulation package \cite{eastman2013}. The used forcefield
was \emph{amber99sb-ildn} \cite{lindorff2010} and the used water-model
was \emph{tip3p} \cite{jorgensen1983}. The peptide was simulated in a
cubic box of $2.7 nm$ length including $652$ solvent molecules.
Langevin equations were integrated at $T=300K$ using a time-step $dt$
of $2fs$.  The potential used for umbrella sampling simulations was
$V_{i}(\phi)=k[1+\cos(\phi-\phi_i-\pi)]$ with
$k=200 kJ/mol$. Umbrellas were placed at a spacing of
$\phi_{i}-\phi_{i+1}=9^{\circ}$.

\subsubsection{Analysis in $\phi$ and $\psi$ dihedral angle space}
We show the convergence of the largest relaxation timescale and
validate the MSM constructed at a lagtime of $\tau=6ps$ via a
Chapman-Kolmogorov test in
\autoref{fig:alanine_validation_phipsi}. Convergence of the largest
relaxation time indicates that the slow eigenfunctions of the
associated dynamical operator are well approximated by the discrete
MSM. The Chapman-Kolmogorov-test explicitly checks the Markov
assumption comparing self transition probabilities computed from the
MSM, parametrized at lagtime $\tau$, with direct estimates from the
data at larger lagtimes, $n\tau$. A thorough discussion of MSM
validation can be found in \cite{prinz2011}.

\begin{figure}
  \includegraphics[width=\columnwidth]{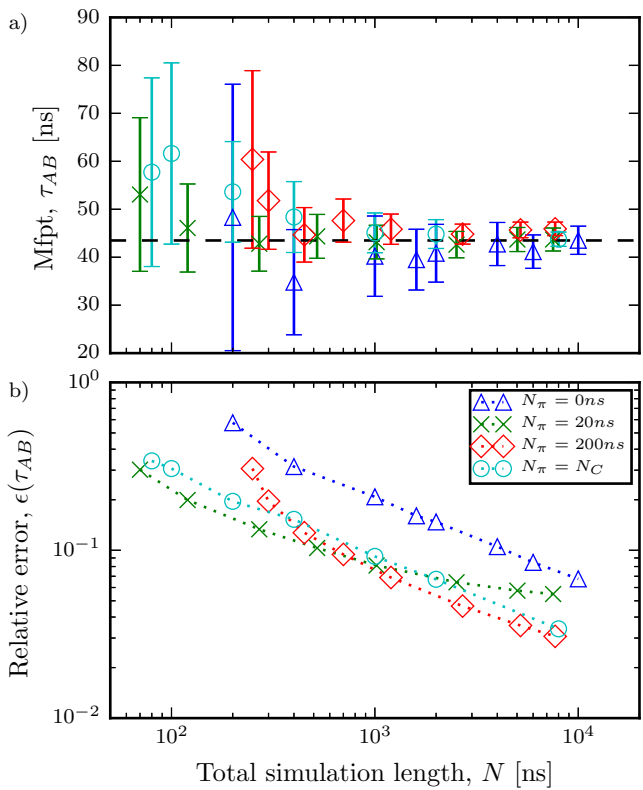}
  \caption{Mean and standard error of mean first passage time (mfpt)
    $\tau_{AB}$, total simulation effort $N$, for alanine-dipeptide
    MSM on the $\phi, \psi$ dihedral angles. The mean first-passage
    time $\tau_{AB}$ of the $C_5$ to $\alpha_{L}$ transition is used
    as an observable for a rare-event process. a) Convergence of the
    mean value is shown for a small number of long chains starting in
    the $C_5$ region (blue), an ensemble of short chains starting in
    the $\alpha_{L}$ region combined with different amounts of
    umbrella sampling simulations (green, red, light-blue). The
    correct value of the $C_{5}$ to $\alpha_{L}$ transition,
    $\tau_{AB}=43 ns$ can be obtained already for a total simulation
    effort of $N=70ns$ when short 'downhill' simulations are used in
    combination with umbrella sampling data. b) The standard error
    shows almost one order of magnitude speedup when estimating the
    kinetic characteristic of a rare-event $\tau_{AB}$ using short
    trajectories in combination with umbrella sampling simulations
    compared to using long trajectories and no additional information
    about the stationary vector.}
  \label{fig:alanine_tAB_phipsi}
\end{figure}

In \autoref{fig:alanine_tAB_phipsi} we show the estimate of the mean
first-passage time $\tau_{AB}$ between the $C_5$ and the $\alpha_{L}$
region together with the corresponding standard error
$\epsilon(\tau_{AB})$ for different values of the total simulation
effort $N$. The simulation setup is similar to the one described for
the double-well potential in the previous section. Instead of starting
short trajectories directly on the barrier we start them from the
meta-stable $\alpha_{L}$ region. \autoref{fig:alanine_tAB_phipsi}
shows that combining umbrella sampling data and short trajectories
relaxing from a meta-stable region with low probability (high
free-energy) towards a meta-stable state with high probability (low
free-energy) is able to estimate the reference value, $\tau_{AB}=43ns$
for the $C_5$ to $\alpha_{L}$ transition with a total simulation
effort of $70ns$ if short 'downhill' trajectories are used in
combination with umbrella sampling data. Utilizing information about
the equilibrium distribution in combination with short simulations
that do not have to sample the rare event is able to achieve a
standard error with almost an order of magnitude less simulation
effort compared to an ensemble of long trajectories. The observed
8-fold speedup is in good agreement with the expected speedup
given by
\begin{equation*}
  \frac{\tau_{AB}}{L}
\end{equation*}
with $\tau_{AB}=43ns$ the mfpt for the slow ``up-hill'' transition
from $C_5$ to $\alpha_L$, and $L=5ns$ the length of individual short
trajectories.

The present approach of estimating rare-event kinetics is more
powerful than traditional rate theories because quantities that can be
estimated can be much more complex than only rates. As a reversible
Markov model is estimated, full mechanisms, such as the ensemble of
transition pathways from one state to another state can be
computed. To illustrate this we compute the committor probability
function, cf. \autoref{sec:committor}, from $C_5$ to $\alpha_L$ using
both estimates. It is seen that information about the stationary
vector results in nearly the same committor function as one estimated
using an order of magnitude larger simulation effort.

\begin{figure}
  \subfloat[]{
    \includegraphics[width=\columnwidth]{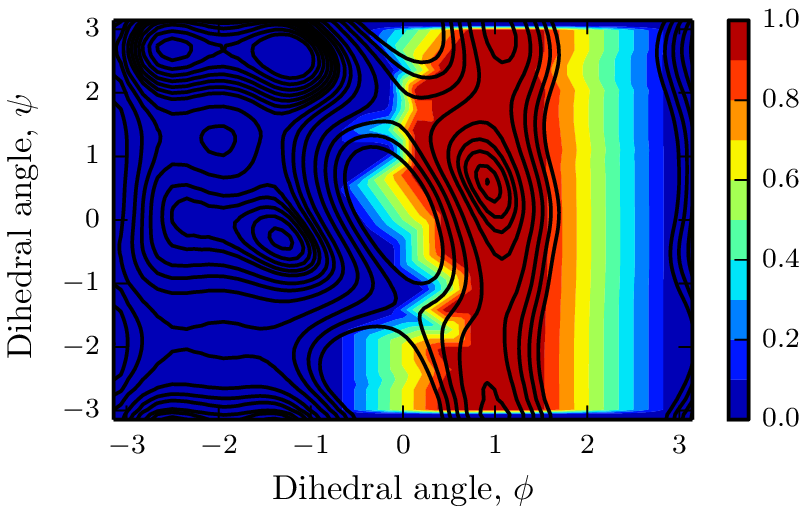}
  }

  \subfloat[]{
    \includegraphics[width=\columnwidth]{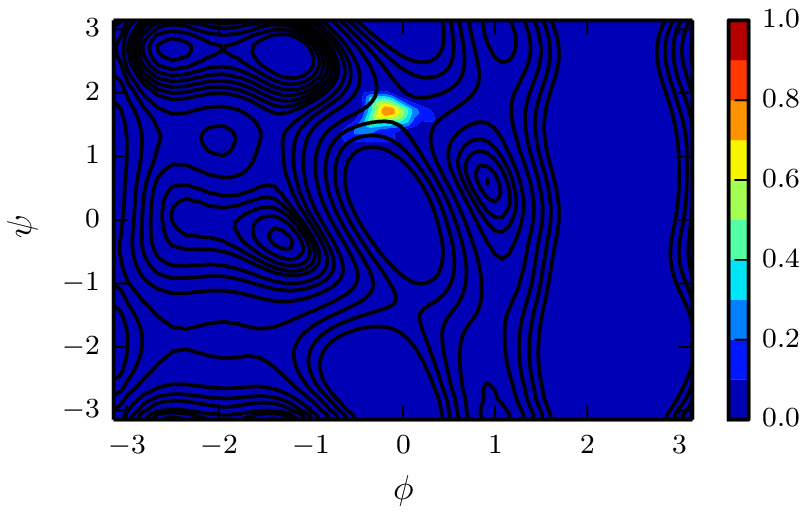}
  }

  \subfloat[]{
    \includegraphics[width=\columnwidth]{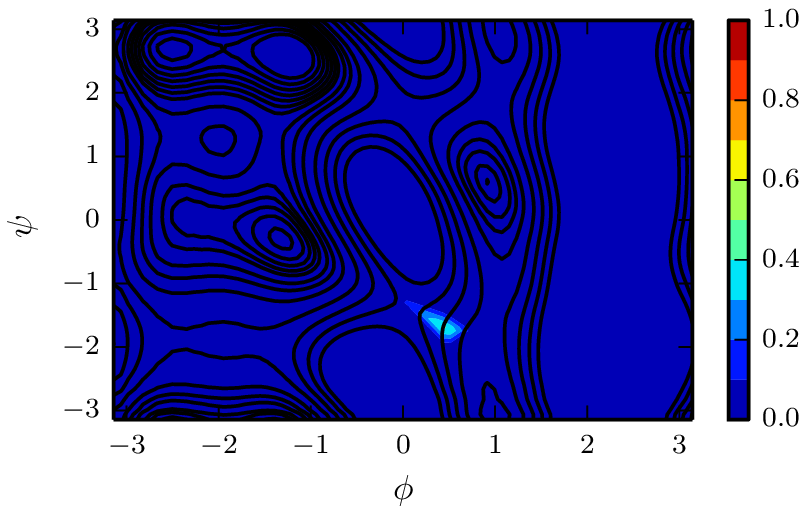}
  }
  \caption{Forward committor $q^{+}(x)$ for transition from $C_5$ to
\label{fig:pmf_alanine_phi}
$\alpha_L$ region. a) shows a non-reversible reference estimate
    for $N=10 \mu s$ of simulation data.  Dark contour lines indicate
    the free energy profile. b) shows the difference between the
    reference estimate and a non-reversible estimate for $N=1 \mu s $
    of simulation data. There is a large error in the transition
    region due to insufficient sampling in the short simulation.  c)
    shows the distance for an estimate using a combination of umbrella
    sampling and standard simulation data with
    $N=N_{\pi}+N_{C}=960ns$. There is no significant error in the
    transition region, the small error close to the second saddle is
    probably due to insufficient sampling of this region by the
    reference simulation.}
\label{fig:committor_diff}
\end{figure}

\subsubsection{Analysis in the $\phi$-coordinate alone}
\begin{figure}
  \centering
  \includegraphics[width=\columnwidth]{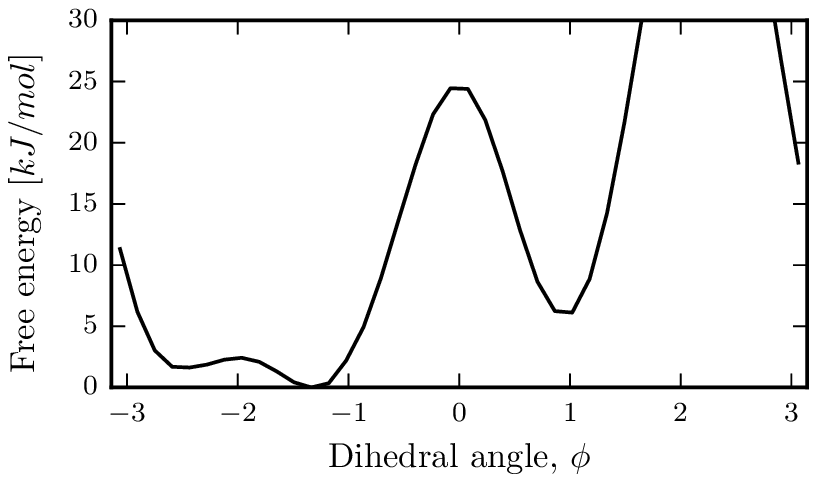}
  \caption{Free energy profile for alanine-dipeptide as a function of
    the $\phi$ dihedral angle. One can identify three meta-stable
    sets. Two low energy (high probability sets) with $\phi < 0$ and a
    single high energy (low probability) set with $\phi>0$.}
  \label{fig:alanine_pmf_phi}
\end{figure}

The presented method can also work if only information about the
slowest degree of freedom is used. In \autoref{fig:alanine_pmf_phi} we
show the free energy profile for the $\phi$ dihedral angle.  An
energetic barrier clearly separates the low free energy region,
$\phi<0$ from the high free energy region, $\phi>0$. Crossing events
from $\phi<0$ to $\phi>0$ are rare leading to a sampling problem if
kinetic quantities associated with barrier-crossings need to
be estimated.
Again we show convergence of the largest relaxation time-scale, $t_2$
and a Chapman-Kolmogorov test for a MSM estimated at lagtime
$\tau=15ps$, \autoref{fig:alanine_validation_phi}.

\begin{figure}
  \includegraphics[width=\columnwidth]{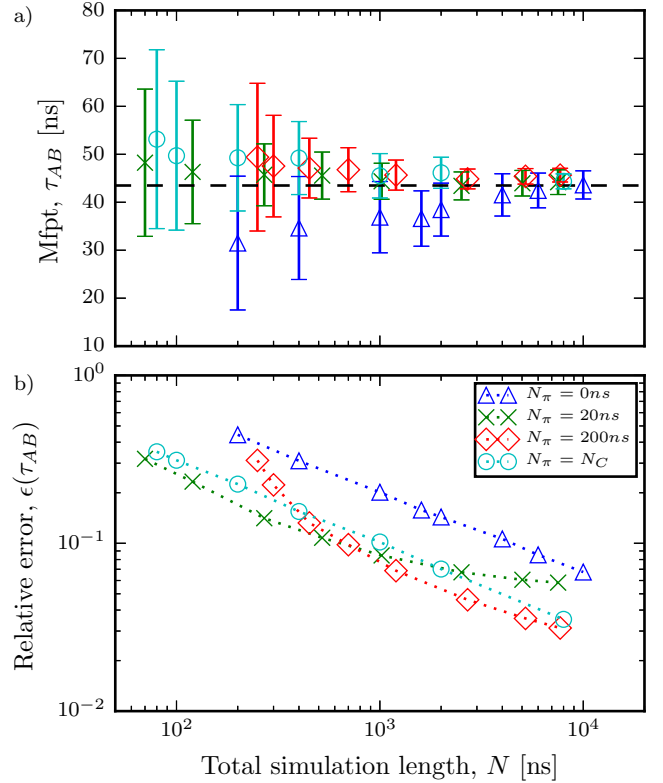}
  \caption{Mean and standard error of mean first passage time (mfpt)
    $\tau_{AB}$, total simulation effort $N$, for alanine-dipeptide
    MSM on the $\phi$ dihedral angle alone. The mean first-passage
    time $\tau_{AB}$ of the transition from the low free energy
    region, $A=\{\phi | -162^{\circ}<\phi<-54^{\circ}\}$, to the high
    free-energy region, $B=\{\phi | 36^{\circ}<\phi<72^{\circ}\}$, is
    used as an observable for a rare-event process. a) Convergence of
    the mean value is shown for a small number of long chains starting
    in the $A$ region (blue), an ensemble of short chains starting in
    the $B$ region combined with different amounts of umbrella
    sampling simulations (green, red, light-blue). The correct value,
    $\tau_{AB}=43ns$, for the $C_5$ to $\alpha_L$ transition can be
    obtained even if no information about the $\psi$ dihedral angle is
    used in the construction of the MSM. b) The standard error shows
    almost one order of magnitude speedup when estimating the kinetic
    characteristic of a rare-event $\tau_{AB}$ using short
    trajectories in combination with umbrella sampling simulations
    compared to using long trajectories and no additional information
    about the stationary vector.}
  \label{fig:alanine_tAB_phi}
\end{figure}

In \autoref{fig:alanine_tAB_phi} we show that the correct mean first
passage time for the $C_5$ to $\alpha_{L}$ transition can also be
recovered from the MSM of the $\phi$ angle alone. This demonstrates
that the presented method is robust with respect to the choice of
microstates. Choosing a slightly larger lagtime $\tau=15ps$ for the
$\phi$ MSM allowed to recover the correct mean first passage times
despite the fact that information about the $\psi$ dihedral angle was
completely neglected. The MSM for $\phi$ dihedral angle is still a
good approximation to the true kinetics if the discretization and the
lagtime are suitably matched. A thorough discussion of approximation
errors for MSMs can be found in \cite{prinz2011,SarichNoeSchuette2010}.

\subsection{Vesicle model}
As a final example we consider the diffusive motion of a colloid
that can reversibly attach to a surface via $m=0,\dots,M$ tethers.
A biological example of such a system is a neuronal vesicle that
can attach to a plasma membrane by SNARE protein complexes. 
The diffusion in the solvent is free but the attachment of tethers restricts the location of the
vesicle to a vicinity of the membrane. The restriction is stronger the
more tethers are attached. Attachment of the vesicle to the membrane
is a fast process, but the dissociation from the membrane is an
extremely rare event. We show that the mean first passage time for
dissociation can be reliably estimated despite the fact that a
non-Markovian coordinate, the membrane-vesicle distance, is used.

\begin{figure}
  \centering
  \includegraphics[width=\columnwidth]{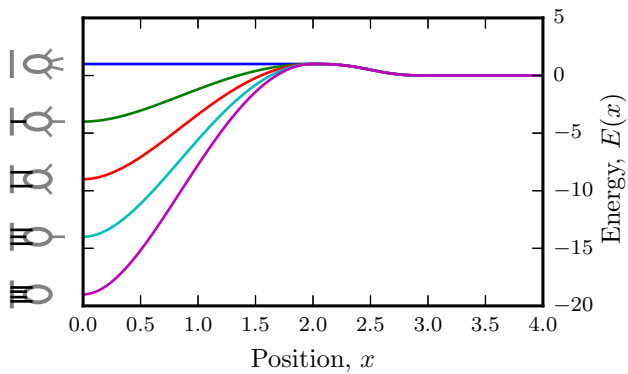}
  \caption{Energy landscape for the different attachment modes,
    $m=0,1,\dots,4$}
  	\label{fig:vesicle_potential}
\end{figure}

\autoref{fig:vesicle_potential} shows the energy for the different
vesicle attachment modes. For $m>0$ attachment of the vesicle to the
membrane is governed by a harmonic potential close to the
membrane. For $x>2$ all attachment modes are energetically equal
corresponding to a breaking of the $m$ tethers once the distance
between the vesicle and the membrane exceeds a certain threshold. The
association of the vesicle has to overcome a small energetic barrier,
modelling a weak repulsion of the untethered vesicle. 

The state of the vesicle is given by the pair $(x, m)$ where $x$ is
the vesicle membrane distance and $m$ denotes the number of tethers
attached. A discretization of the vesicle membrane distance with
$0=x_1<\dots<x_d=4$ allows us to describe the vesicle dynamics by a
Markov chain on a finite state space with $(M+1)d$ microstates. The
stationary vector of the chain is given as
\begin{equation}
  \label{eq:stationary_vector_vesicle}
  \pi = (\pi^{(0)}(x_1),\dots,\pi^{(M)}(x_d))
\end{equation}
with entries given in terms of the usual Gibbs/Boltzmann distribution,
\begin{equation}
  \label{eq:stationary_vector_element_vesicle}
  \pi^{(m)}(x_i) \propto e^{-E^{(m)}(x_i)}.
\end{equation}
$E^{(m)}(x)$ is the energy of a vesicle at $x$ with $m$ tethers
attached, cf. \autoref{fig:vesicle_potential},
\eqref{eq:vesicle_analytical}.

The transition matrix
$P=(p_{ij})$ for the vesicle dynamics is now constructed as
follows. We encode random walk probabilities in a proposal matrix
$Q=(q_{ij})$. The particle moves from $x_{i}$ to $x_{i-1}$ or
$x_{i+1}$ with probability $1/3$, if the particle remains at its
current position $x_i$ it can attach, $m \to m+1$, or detach, $m \to
m-1$, a tether with probability $1/3$ so that the overall proposal-probability
for attachment or detachment is $1/9$. To account for the energetic
differences of the microstates we use the Metropolis-Hastings
acceptance criterion to modulate the proposal probabilities and obtain
the desired transition probabilities via,
\begin{equation}
  \label{eq:transition_matrix_vesicle}
  p_{ij} = \min \{1, \frac{\pi_j q_{ji}}{\pi_i q_{ij}} \} \quad i \neq j.
\end{equation}
Correct normalization is ensured by setting $p_{ii} = 1 - \sum_{j\neq
  i} p_{ij}$. As a result of \eqref{eq:transition_matrix_vesicle} the
constructed transition matrix $P$ automatically fulfills the detailed
balance condition \eqref{eq:defn_detailed_balance} with respect to the
desired stationary vector.

The mean first-passage time for the dissociation of the vesicle is
$\tau_{AB} = 8.56\cdot 10^{9}$, the mean first-passage time for
association, $\tau_{BA} = 1.59 \cdot 10^{3}$, is orders of magnitude
smaller.  The mean first-passage time for dissociation of a vesicle
with the maximum number of tethers attached is $\tau_{AB} = 3.83 \cdot
10^{10}$ so that the system dynamics can not be described in terms of
the subspace with $m=4$ tethers. This indicates that the dissociation
kinetics is effectively non-Markovian along the $x$-coordinate.

\begin{figure}
  \centering
  \includegraphics[width=\columnwidth]{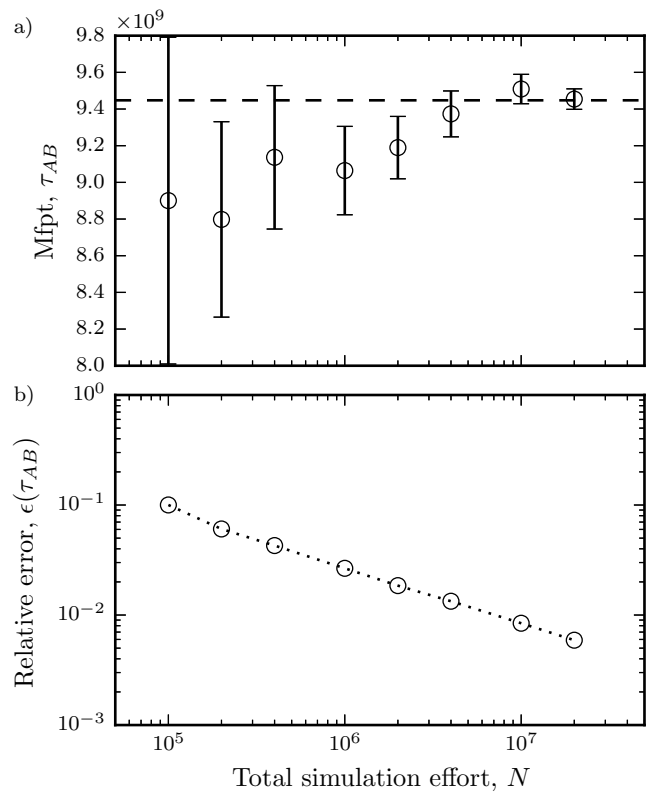}
  \caption{Mean and standard error of mean first-passage time of
    vesicle dissociation on a coarse-grained non-Markovian state
    space. a) Convergence of the mean value, b) standard
    error. Estimates are obtained for an ensemble of association
    trajectories starting in the high energy region and relaxing
    towards the low energy region in combination with the
    coarse-grained stationary probabilities. The dissociation time
    can be estimated orders of magnitude before a single dissociation
    event would have been observed.}
  \label{fig:mfpt_vesicle}
\end{figure}

The dissociation time $\tau_{AB}$ can reliably be estimated even if no
information about the mode of attachment is available. If only
information about the position of the vesicle is available then the
state-space of the $(M+1)d$ distinct microstates is coarse-grained
into $d$ distinct sets each containing $(M+1)$ microstates
corresponding to the $M+1$ possible tethering modes at position
$x$. The coarse grained stationary vector $\tilde{\pi}$ is obtained by
summing the full stationary vector $\pi$ over all possible tethering
modes. If short association trajectories starting in the region $x>2$
are combined with the coarse-grained stationary vector the
dissociation time can again be estimated orders of magnitude before a
single dissociation event would on average be observed despite the
fact that the MSM is built on a coordinate that is inherently
non-Markovian. In \autoref{fig:mfpt_vesicle} we show mean and standard
error for the mfpt of vesicle dissociation for a MSM build at a
lagtime of $\tau=60$ with $d=40$ microstates.

In \autoref{fig:vesicle_validation} we again show convergence of the
largest relaxation time and the Chapman-Kolmogorov test for an MSM
constructed at lagtime $\tau=60$. The MSM is estimated solely from
short association trajectories starting in the high energy region
using the coarse grained stationary vector $\tilde{\pi}$ to obtain a
reversible maximum likelihood transition matrix from
\eqref{eq:mle_constrained}. The total simulation effort, $N=2 \cdot
10^7$, used to obtain the MSM and perform the validation is again
orders of magnitude smaller than the expected dissociation time.

%% file: discussion.tex
\section{Conclusion}
We have described a principle that allows for the first time to
estimate rare-event kinetics efficiently without having to assume a
simple kinetic model. Our approach is applicable when the kinetic
properties of interest can be computed from a Markov state model
discretization of the system.  Note that this
approach is qualitatively different from assuming a specific rate
theory such as Transition State Theory or Kramers, because MSMs are a
numerical approximation method of the full kinetics and can be made
arbitrarily accurate in the limit of a good state space discretization
\cite{prinz2011} whereas a specific rate model needs to apply by
design and can usually not be self-consistently validated.

The key idea of the presented approach is to use enhanced sampling
methods to obtain reliable estimates of the equilibrium distribution
in combination with direct simulations of the fast downhill
processes. These data are combined rigorously in a reversible Markov
model. Our approach can deliver estimates of kinetic properties,
including rates, passage times, but also complex quantities such as
committor functions and transition path ensembles while achieving
enormous speedups compared to a direct simulation.

We have illustrated our method using two toy models, a explicit-solvent
MD simulation of alanine dipeptide with about 2000 degrees of freedom,
and a model for reversible attachment of a vesicle to a membrane. In
these examples, the kinetics of the rare events could be computed 
using between one and six orders of magnitude less simulation time
than needed with a direct simulation approach that has to wait for the
rare events to happen spontaneously.

In general, the present approach will be efficient whenever the rare-event occurs
between low-probability and high-probability states. A very important
example of this class is computational drug design, where the binding
of the drug compound occurs relatively fast
\cite{BuchFabritiis_PNAS11_Binding}, while the unbinding may be many
orders of magnitude slower. Yet the unbinding kinetics have been shown 
to be critical for drug efficacy \cite{tummino2008}.

We have demonstrated in two applications that the approach can compute
kinetics from non-Markovian projections of the data: By using only the
$\phi$-coordinate in alanine dipeptide, and by using only the distance
coordinate in the vesicle attachment model. A requirement is that the resolved
coordinates are slow compared to the non-resolved coordinates. However, this
requirement is not overly restrictive, as the same requirement applies for
the enhanced sampling simulations, such as umbrella sampling, employed
to obtain an estimate for the stationary distribution. 

While the applications in the present paper have used reversible
Markov model estimates in such a way that the enhanced sampling
simulation and the unbiased ``downhill'' simulations visit the same
state space, the principle explored here can be generalized beyond this
case. States visited only in one but not in the other simulation can
be modelled by appropriate uninformative priors on the respective
variables, e.g. uniform prior in the equilibrium distributions of
states not visited in an umbrella sampling simulation. 

A general framework to reconcile direct MD and enhanced MD simulations
is the transition-based reweighting analysis method (TRAM) framework
\cite{WuNoe2014, WuMeyRostaNoe2014, MeyWuNoe2014}.  In order to apply
the TRAM framework to the current setting, a hybrid TRAM method must
be developed that can mix kinetic simulations (with an estimation lag
time tau), and simulations that contains trajectories shorter than
tau, such as those used in umbrella sampling or REMD.

The present inference principle can be exploited in an adaptive sampling
framework \cite{BowmanEnsignPande_JCTC2010_AdaptiveSampling,
  doerr2014} to optimally distribute the computational effort between
enhanced sampling and unbiased molecular dynamics simulations.

%% file: acknowledge.tex
\begin{acknowledgements}
  We are grateful to Feliks N\"{u}ske for stimulating
  discussions. This work was funded by the Deutsche
  Forschungsgemeinschaft (DFG) grant NO825/3-1 (BTS) and a
  European Research Council (ERC) starting grant pcCell (FN).
\end{acknowledgements}

%% file: appendix.tex
\appendix

\section{The transition kernel for the Euler-method}
\label{sec:transition_kernel}
The solution of \eqref{eq:Brownian_dynamics} with initial position
$X_0=x_0$ on $[0, T]$ is usually carried out by choosing a regular
discretization of the time interval
\begin{equation*}
  0=t_0<t_1<\dotsb<t_N=T.
\end{equation*}
with $\Delta t=t_{k}-t_{k-1}$ for all $k=1,\dotsc,N$. The evolution of
the stochastic process is then approximated by the following
time-stepping scheme
\begin{equation}
  \label{eq:Euler_scheme}
  X_{t+\Delta t}=X_t-\nabla V(X_t)\Delta t + \sqrt{2 \beta^{-1}} \eta.
\end{equation}
with $X_0=x_0$ and $\eta$ being a $\mathcal{N}(0, \Delta t)$
distributed random variable. The time-stepping scheme
\eqref{eq:Euler_scheme} is known as \emph{Euler method} or
\emph{Euler-Maruyama method}, \cite{kloeden1992}.

For this simple time-stepping scheme the transition kernel of the
resulting Markov chain is given by
\begin{equation}
  \label{eq:transition_kernel}
  p_{\Delta t}(x, y)=\frac{1}{\sqrt{2\pi}\Delta t 2/\beta} \exp \left(-\frac{(y-x+\nabla V(x)\Delta t)^2}{2(\sqrt{\Delta t}\sqrt{2/\beta})^2} \right),
\end{equation}
with  $x=X_t$  and  $y=X_{t+\Delta  t}$.  $p_{\Delta t}(x,  y)$  is  a
Gaussian  distribution  with mean  $\mu=x-\nabla  V(x)  \Delta t$  and
variance $\sigma^2=2 \Delta t / \beta$.

The transition probability $P_{\Delta t}(B|A)$
between two sets $A$, $B$ can be computed from
\begin{equation}
  \label{eq:transition_probability_sets}
  P_{\Delta t}(B|A)=\frac{\int_{A}\mathrm{d}x \, \pi(x) \int_{B} \mathrm{d}y \, p_{\Delta t}(x, y) }{\int_{A} \mathrm{d}x \, \pi(x) }.
\end{equation}

Choosing a $L$ such that $p_{\Delta t}(x, y)$ is effectively zero
outside of $[-L, L]$ we pick a spatial discretization
\begin{equation}
  \label{eq:spatial_discretization}
  -L=x_0<x_1<\dotsc<x_N=L
\end{equation}
with a regular spacing $\Delta x= x_{k}-x_{k-1}$ for $k=1,\dotsc,N$
such that $p_{\Delta t}(x, y)$ and $\pi(x)$ are approximately constant
on sub-intervals $S_i=(x_{k}, x_{k+1}]$. In this case we have
\begin{equation*}
  \int_{x_{i}}^{x_{i+1}}\mathrm{d}x \, \mu(x) \approx \mu(x_{k})\Delta x
\end{equation*}
and
\begin{equation*}
  \int_{x_{i}}^{x_{i+1}} \mathrm{d}x \, \mu(x) \int_{x_{j}}^{x_{j+1}} \mathrm{d}y \, p(x, y) \approx \mu(x_i)p(x_{i}, x_{j})(\Delta x)^2.
\end{equation*}
We can approximate the matrix elements $p_{ij}=P(S_j|S_i)$ as
\begin{equation*}
  p_{ij} \approx p(x_i, x_j) \Delta x.
\end{equation*} 
and compute spectral properties from the matrix $(p_{ij})$ using
standard eigenvalue solvers.

\section{Mean first-passage times between meta-stable regions}
\label{sec:mfpt}
The covered material can be found in many introductory books to
stochastic processes, cf. \cite{hoel1986}.  

For a stochastic process $(X_t)$ on a state space $\Omega$ the first
hitting time $T_{B}$ of a set $B \subseteq \Omega$ is defined as
\begin{equation}
  \label{eqn:first_hitting_time}
  T_{B}= \inf \{ t \geq 0 | X_{t} \in B\}.
\end{equation}
The mean first passage time $\tau_{x, B}$ to the set $B$ starting in
state $x \in \Omega$ is the following expectation value
\begin{equation}
  \label{eqn:mfpt_state_to_set}
  \tau_{x, B}=\mathbb{E}_{x}(T_{B}).
 \end{equation}
 For a Markov chain on a finite state space $\Omega=\{1,\dotsc,n\}$
 with transition matrix $(p_{x, y})$ the mean first-passage time can
 be computed from the following system of equations,
\begin{equation}
  \label{eqn:mfpt_system}
  \tau_{x, B}=\left \{ \begin{array}{cc} 0 & x \in B \\ 1 + \sum_{y \in \Omega} p_{x, y} \tau_{y, B} & x \notin B \end{array} \right .
\end{equation}

Assuming that the chain has stationary vector $(\mu_{x})$ we define
the mean first-passage time $\tau_{A,B}$ from set $A$ to set $B$ as
the $\mu$-weighted average of all mean first-passage times to $B$
when starting in a state $x \in A$,
\begin{equation}
  \label{eqn:mfpt_set_to_set}
  \tau_{A, B}=\sum_{x \in A} \mu_{x} \tau_{x, B}.
\end{equation}

Computing the mean first-passage time between two sets for a Markov
chain on a finite state space with given transition matrix thus
amounts to finding the stationary vector together with the solution of
a linear system of equations - both of which can be achieved using
standard numerical linear algebra libraries.

\section{Committor functions}
\label{sec:committor}
Committor functions have been introduced in the context of Transition
Path Theory \cite{vanden2006} and are a central object for the
characterisation of transition processes between two meta-stable sets.

Let $(X_t)$ again be a stochastic process on a state space $\Omega$
and let $A, B \subseteq \Omega$ be two meta-stable sets. The forward
committor $q^{(+)}(x)$ is the probability that the process starting in
$x$ will reach the set $B$ first, rather than the set $A$,
\begin{equation}
  \label{eq:fwd_committor}
  q^{(+)}(x)=\mathbb{P}_x(T_{A} < T_{B}).
\end{equation}
Again $T_{S}$ denotes the first hitting time of a set $S$.

For a Markov chain on a finite state space with transition matrix
$P$ the forward committor solves the following boundary
value problem \cite{metzner2009},
\begin{equation}
  \label{eq:fwd_committor_bvp}
  \begin{array}{rl}
    \sum_{j} l_{ij}q^{(+)}_{j}=0 & i \in X \ (A \cup B) \\
    q_{i}^{(+)} = 0 & i \in A \\
    q_{i}^{(+)} = 1 & i \in B 
    \end{array} .
\end{equation}
$L=P-I$ is the corresponding generator matrix of the Markov chain. 

Computing the committor for a finite state space again amounts to solving a
linear system of equations.

\begin{widetext}
\section{Vesicle potential}

\begin{equation}
  \label{eq:vesicle_analytical}
  E^{(m)}(x) = \left \{
    \begin{array}{cc}
      1 + m (-5 + 5 x^2 -2.5 x^3 + 0.3125 x^4) & 0 \leq x < 2 \\
      1 + 8 (x-2)^2 - 8(x-2)^3 & 2 \leq x < 2.5 \\
      0.5 - 8 (x-2.5)^2 + 8 (x-2.5)^3 & 2.5 \leq x < 3 \\
      0 & 3 \leq x < 4
    \end{array} 
    \right .
\end{equation}
\section{MSM validation}

\end{widetext}

\begin{figure}
  \centering
  \subfloat[]{
    \includegraphics[width=\columnwidth]{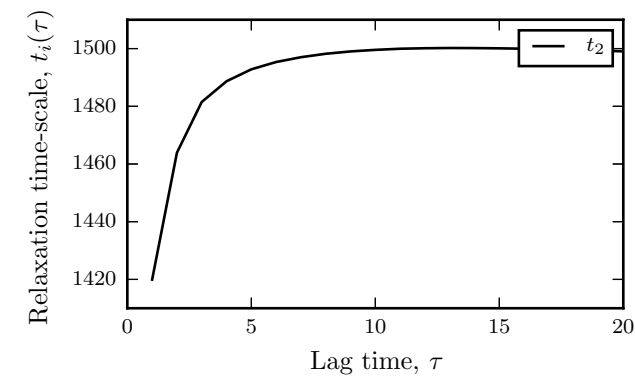}
  }

  \subfloat[]{
    \includegraphics[width=\columnwidth]{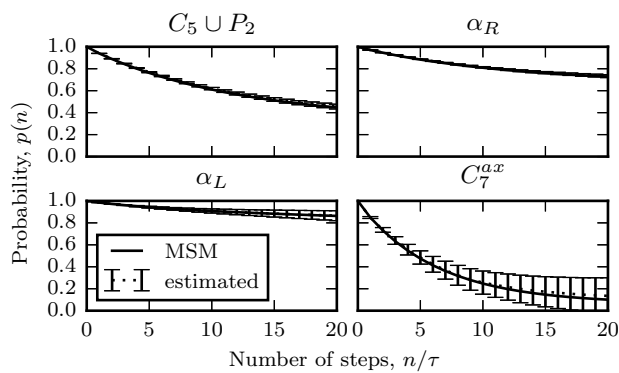}
  }
  \caption{a) Implied timescale test. Convergence of the largest
    relaxation time-scale, $t_2$, indicates a good Markov model fit,
    i.e. the slow eigenfunction of the associated dynamical operator
    are well approximated. b) The Chapman-Kolmogorov test validates
    the Markov assumption by comparing the evolution of
    self-transition probabilities predicted by the MSM parametrized at
    lagtime $\tau$ with direct estimates from the data at larger
    lagtimes $n \tau$.}
  \label{fig:alanine_validation_phipsi}
\end{figure}

\begin{figure}
  \centering
  \subfloat[]{
    \includegraphics[width=\columnwidth]{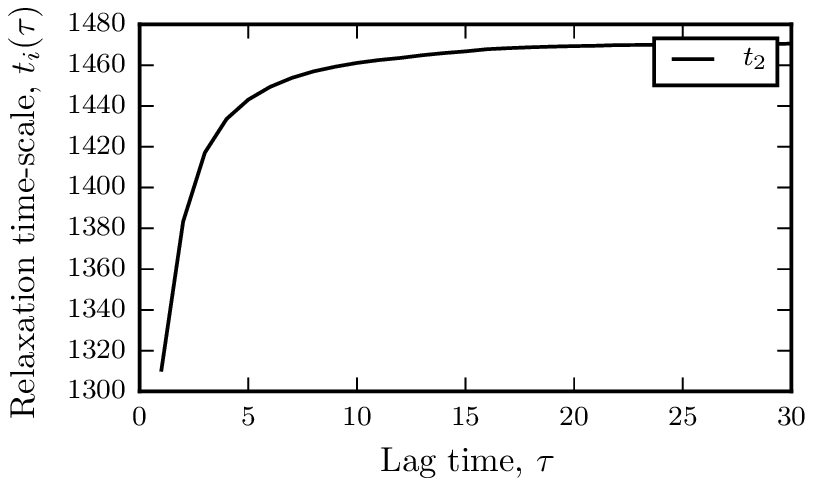}
  }

  \subfloat[]{
    \includegraphics[width=\columnwidth]{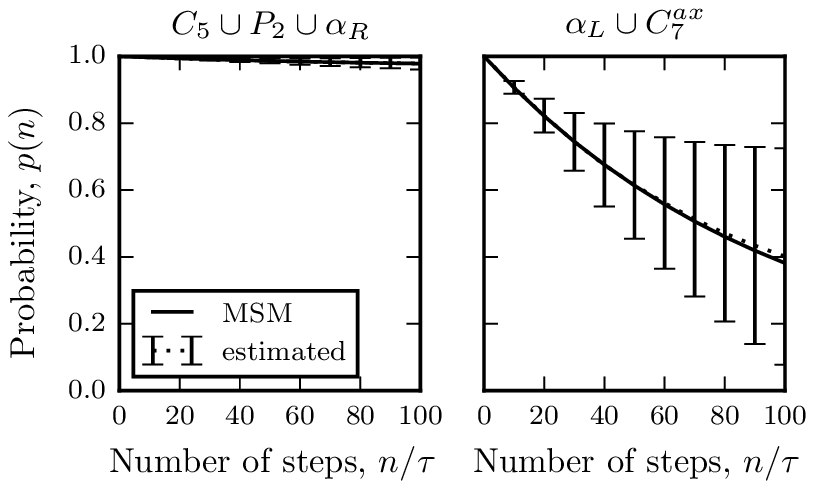}
  }
  \caption{a) Implied timescale test. Convergence of the largest
    relaxation time-scale, $t_2$, indicates a good Markov model fit,
    i.e. the slow eigenfunction of the associated dynamical operator
    are well approximated. b) The Chapman-Kolmogorov test validates
    the Markov assumption by comparing the evolution of
    self-transition probabilities predicted by the MSM parametrized at
    lagtime $\tau$ with direct estimates from the data at larger
    lagtimes $n \tau$.}
  \label{fig:alanine_validation_phi}
\end{figure}

\begin{figure}
  \centering
  \subfloat[]{
    \includegraphics[width=\columnwidth]{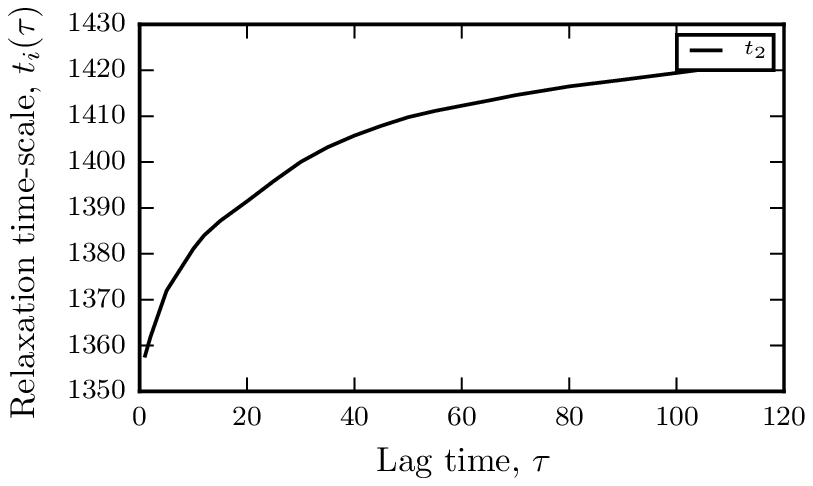}
  }

  \subfloat[]{
    \includegraphics[width=\columnwidth]{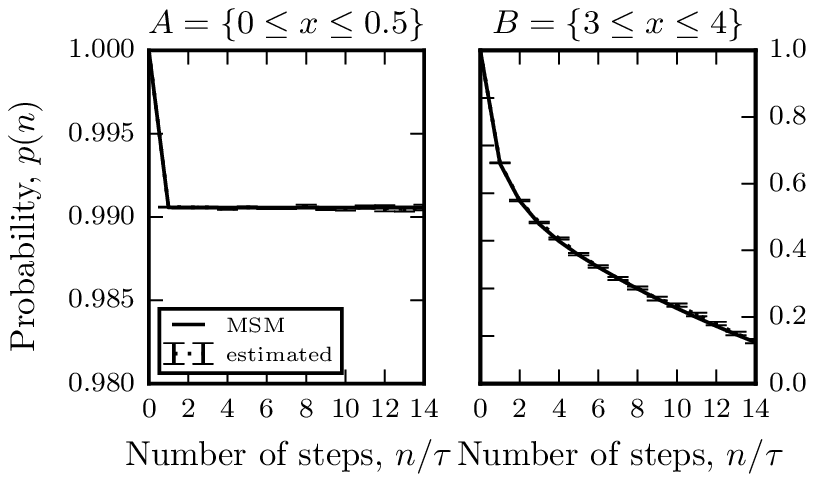}
  }
  \caption{a) Implied timescale test. Convergence of the largest
    relaxation time-scale, $t_2$, indicates a good Markov model fit,
    i.e. the slow eigenfunction of the associated dynamical operator
    are well approximated. b) The Chapman-Kolmogorov test validates
    the Markov assumption by comparing the evolution of
    self-transition probabilities predicted by the MSM parametrized at
    lagtime $\tau$ with direct estimates from the data at larger
    lagtimes $n \tau$. Values were obtained from an ensemble of short
    trajectories starting in the high energy region utilizing the
    stationary vector in the estimation of the MLE transition matrix,
    cf. \eqref{eq:mle_constrained}.}
  \label{fig:vesicle_validation}
\end{figure}

%% file: publication.bbl
\begin{thebibliography}{68}%
\makeatletter
\providecommand \@ifxundefined [1]{%
 \@ifx{#1\undefined}
}%
\providecommand \@ifnum [1]{%
 \ifnum #1\expandafter \@firstoftwo
 \else \expandafter \@secondoftwo
 \fi
}%
\providecommand \@ifx [1]{%
 \ifx #1\expandafter \@firstoftwo
 \else \expandafter \@secondoftwo
 \fi
}%
\providecommand \natexlab [1]{#1}%
\providecommand \enquote  [1]{``#1''}%
\providecommand \bibnamefont  [1]{#1}%
\providecommand \bibfnamefont [1]{#1}%
\providecommand \citenamefont [1]{#1}%
\providecommand \href@noop [0]{\@secondoftwo}%
\providecommand \href [0]{\begingroup \@sanitize@url \@href}%
\providecommand \@href[1]{\@@startlink{#1}\@@href}%
\providecommand \@@href[1]{\endgroup#1\@@endlink}%
\providecommand \@sanitize@url [0]{\catcode `\\12\catcode `\$12\catcode
  `\&12\catcode `\#12\catcode `\^12\catcode `\_12\catcode `\%12\relax}%
\providecommand \@@startlink[1]{}%
\providecommand \@@endlink[0]{}%
\providecommand \url  [0]{\begingroup\@sanitize@url \@url }%
\providecommand \@url [1]{\endgroup\@href {#1}{\urlprefix }}%
\providecommand \urlprefix  [0]{URL }%
\providecommand \Eprint [0]{\href }%
\providecommand \doibase [0]{http://dx.doi.org/}%
\providecommand \selectlanguage [0]{\@gobble}%
\providecommand \bibinfo  [0]{\@secondoftwo}%
\providecommand \bibfield  [0]{\@secondoftwo}%
\providecommand \translation [1]{[#1]}%
\providecommand \BibitemOpen [0]{}%
\providecommand \bibitemStop [0]{}%
\providecommand \bibitemNoStop [0]{.\EOS\space}%
\providecommand \EOS [0]{\spacefactor3000\relax}%
\providecommand \BibitemShut  [1]{\csname bibitem#1\endcsname}%
\let\auto@bib@innerbib\@empty
%</preamble>
\bibitem [{\citenamefont {Sch{\"u}tte}\ \emph {et~al.}(1999)\citenamefont
  {Sch{\"u}tte}, \citenamefont {Fischer}, \citenamefont {Huisinga},\ and\
  \citenamefont {Deuflhard}}]{huisinga1999}%
  \BibitemOpen
  \bibfield  {author} {\bibinfo {author} {\bibfnamefont {C.}~\bibnamefont
  {Sch{\"u}tte}}, \bibinfo {author} {\bibfnamefont {A.}~\bibnamefont
  {Fischer}}, \bibinfo {author} {\bibfnamefont {W.}~\bibnamefont {Huisinga}}, \
  and\ \bibinfo {author} {\bibfnamefont {P.}~\bibnamefont {Deuflhard}},\
  }\href@noop {} {\bibfield  {journal} {\bibinfo  {journal} {J. Comp. Phys.}\
  }\textbf {\bibinfo {volume} {151}},\ \bibinfo {pages} {146} (\bibinfo {year}
  {1999})}\BibitemShut {NoStop}%
\bibitem [{\citenamefont {Swope}\ \emph {et~al.}(2004)\citenamefont {Swope},
  \citenamefont {Pitera},\ and\ \citenamefont {Suits}}]{swope2004}%
  \BibitemOpen
  \bibfield  {author} {\bibinfo {author} {\bibfnamefont {W.~C.}\ \bibnamefont
  {Swope}}, \bibinfo {author} {\bibfnamefont {J.~W.}\ \bibnamefont {Pitera}}, \
  and\ \bibinfo {author} {\bibfnamefont {F.}~\bibnamefont {Suits}},\
  }\href@noop {} {\bibfield  {journal} {\bibinfo  {journal} {J. Phys. Chem. B}\
  }\textbf {\bibinfo {volume} {108}},\ \bibinfo {pages} {6571} (\bibinfo {year}
  {2004})}\BibitemShut {NoStop}%
\bibitem [{\citenamefont {Singhal}\ \emph {et~al.}(2004)\citenamefont
  {Singhal}, \citenamefont {Snow},\ and\ \citenamefont {Pande}}]{singhal2004}%
  \BibitemOpen
  \bibfield  {author} {\bibinfo {author} {\bibfnamefont {N.}~\bibnamefont
  {Singhal}}, \bibinfo {author} {\bibfnamefont {C.~D.}\ \bibnamefont {Snow}}, \
  and\ \bibinfo {author} {\bibfnamefont {V.~S.}\ \bibnamefont {Pande}},\
  }\href@noop {} {\bibfield  {journal} {\bibinfo  {journal} {J. Chem. Phys.}\
  }\textbf {\bibinfo {volume} {121}},\ \bibinfo {pages} {415} (\bibinfo {year}
  {2004})}\BibitemShut {NoStop}%
\bibitem [{\citenamefont {Chodera}\ \emph {et~al.}(2007)\citenamefont
  {Chodera}, \citenamefont {Dill}, \citenamefont {Singhal}, \citenamefont
  {Pande}, \citenamefont {Swope},\ and\ \citenamefont
  {Pitera}}]{ChoderaEtAl_JCP07}%
  \BibitemOpen
  \bibfield  {author} {\bibinfo {author} {\bibfnamefont {J.~D.}\ \bibnamefont
  {Chodera}}, \bibinfo {author} {\bibfnamefont {K.~A.}\ \bibnamefont {Dill}},
  \bibinfo {author} {\bibfnamefont {N.}~\bibnamefont {Singhal}}, \bibinfo
  {author} {\bibfnamefont {V.~S.}\ \bibnamefont {Pande}}, \bibinfo {author}
  {\bibfnamefont {W.~C.}\ \bibnamefont {Swope}}, \ and\ \bibinfo {author}
  {\bibfnamefont {J.~W.}\ \bibnamefont {Pitera}},\ }\href@noop {} {\bibfield
  {journal} {\bibinfo  {journal} {J. Chem. Phys.}\ }\textbf {\bibinfo {volume}
  {126}},\ \bibinfo {pages} {155101} (\bibinfo {year} {2007})}\BibitemShut
  {NoStop}%
\bibitem [{\citenamefont {No\'{e}}\ \emph {et~al.}(2007)\citenamefont
  {No\'{e}}, \citenamefont {Horenko}, \citenamefont {Sch\"{u}tte},\ and\
  \citenamefont {Smith}}]{NoeHorenkeSchutteSmith_JCP07_Metastability}%
  \BibitemOpen
  \bibfield  {author} {\bibinfo {author} {\bibfnamefont {F.}~\bibnamefont
  {No\'{e}}}, \bibinfo {author} {\bibfnamefont {I.}~\bibnamefont {Horenko}},
  \bibinfo {author} {\bibfnamefont {C.}~\bibnamefont {Sch\"{u}tte}}, \ and\
  \bibinfo {author} {\bibfnamefont {J.~C.}\ \bibnamefont {Smith}},\ }\href@noop
  {} {\bibfield  {journal} {\bibinfo  {journal} {J. Chem. Phys.}\ }\textbf
  {\bibinfo {volume} {126}},\ \bibinfo {pages} {155102} (\bibinfo {year}
  {2007})}\BibitemShut {NoStop}%
\bibitem [{\citenamefont {Pan}\ and\ \citenamefont {Roux}(2008)}]{pan2008}%
  \BibitemOpen
  \bibfield  {author} {\bibinfo {author} {\bibfnamefont {A.}~\bibnamefont
  {Pan}}\ and\ \bibinfo {author} {\bibfnamefont {B.}~\bibnamefont {Roux}},\
  }\href@noop {} {\bibfield  {journal} {\bibinfo  {journal} {J. Chem. Phys.}\
  }\textbf {\bibinfo {volume} {129}} (\bibinfo {year} {2008})}\BibitemShut
  {NoStop}%
\bibitem [{\citenamefont {Buchete}\ and\ \citenamefont
  {Hummer}(2008)}]{BucheteHummer_JPCB08}%
  \BibitemOpen
  \bibfield  {author} {\bibinfo {author} {\bibfnamefont {N.~V.}\ \bibnamefont
  {Buchete}}\ and\ \bibinfo {author} {\bibfnamefont {G.}~\bibnamefont
  {Hummer}},\ }\href@noop {} {\bibfield  {journal} {\bibinfo  {journal} {J.
  Phys. Chem. B}\ }\textbf {\bibinfo {volume} {112}},\ \bibinfo {pages} {6057}
  (\bibinfo {year} {2008})}\BibitemShut {NoStop}%
\bibitem [{\citenamefont {Prinz}\ \emph {et~al.}(2011)\citenamefont {Prinz},
  \citenamefont {Wu}, \citenamefont {Sarich}, \citenamefont {Keller},
  \citenamefont {Senne}, \citenamefont {Held}, \citenamefont {Chodera},
  \citenamefont {Sch{\"u}tte},\ and\ \citenamefont {No{\'e}}}]{prinz2011}%
  \BibitemOpen
  \bibfield  {author} {\bibinfo {author} {\bibfnamefont {J.-H.}\ \bibnamefont
  {Prinz}}, \bibinfo {author} {\bibfnamefont {H.}~\bibnamefont {Wu}}, \bibinfo
  {author} {\bibfnamefont {M.}~\bibnamefont {Sarich}}, \bibinfo {author}
  {\bibfnamefont {B.}~\bibnamefont {Keller}}, \bibinfo {author} {\bibfnamefont
  {M.}~\bibnamefont {Senne}}, \bibinfo {author} {\bibfnamefont
  {M.}~\bibnamefont {Held}}, \bibinfo {author} {\bibfnamefont {J.~D.}\
  \bibnamefont {Chodera}}, \bibinfo {author} {\bibfnamefont {C.}~\bibnamefont
  {Sch{\"u}tte}}, \ and\ \bibinfo {author} {\bibfnamefont {F.}~\bibnamefont
  {No{\'e}}},\ }\href@noop {} {\bibfield  {journal} {\bibinfo  {journal} {J.
  Chem. Phys.}\ }\textbf {\bibinfo {volume} {134}},\ \bibinfo {pages} {174105}
  (\bibinfo {year} {2011})}\BibitemShut {NoStop}%
\bibitem [{\citenamefont {Senne}\ \emph {et~al.}(2012)\citenamefont {Senne},
  \citenamefont {Trendelkamp-Schroer}, \citenamefont {Mey}, \citenamefont
  {Sch\"{u}tte},\ and\ \citenamefont {No\'{e}}}]{senne2012}%
  \BibitemOpen
  \bibfield  {author} {\bibinfo {author} {\bibfnamefont {M.}~\bibnamefont
  {Senne}}, \bibinfo {author} {\bibfnamefont {B.}~\bibnamefont
  {Trendelkamp-Schroer}}, \bibinfo {author} {\bibfnamefont {A.~S.}\
  \bibnamefont {Mey}}, \bibinfo {author} {\bibfnamefont {C.}~\bibnamefont
  {Sch\"{u}tte}}, \ and\ \bibinfo {author} {\bibfnamefont {F.}~\bibnamefont
  {No\'{e}}},\ }\href@noop {} {\bibfield  {journal} {\bibinfo  {journal} {J.
  Chem. Theory Comput.}\ }\textbf {\bibinfo {volume} {8}},\ \bibinfo {pages}
  {2223} (\bibinfo {year} {2012})}\BibitemShut {NoStop}%
\bibitem [{\citenamefont {Beauchamp}\ \emph {et~al.}(2011)\citenamefont
  {Beauchamp}, \citenamefont {Bowman}, \citenamefont {Lane}, \citenamefont
  {Maibaum}, \citenamefont {Haque},\ and\ \citenamefont
  {Pande}}]{BeauchampEtAl_MSMbuilder2}%
  \BibitemOpen
  \bibfield  {author} {\bibinfo {author} {\bibfnamefont {K.~A.}\ \bibnamefont
  {Beauchamp}}, \bibinfo {author} {\bibfnamefont {G.~R.}\ \bibnamefont
  {Bowman}}, \bibinfo {author} {\bibfnamefont {T.~J.}\ \bibnamefont {Lane}},
  \bibinfo {author} {\bibfnamefont {L.}~\bibnamefont {Maibaum}}, \bibinfo
  {author} {\bibfnamefont {I.~S.}\ \bibnamefont {Haque}}, \ and\ \bibinfo
  {author} {\bibfnamefont {V.~S.}\ \bibnamefont {Pande}},\ }\href@noop {}
  {\bibfield  {journal} {\bibinfo  {journal} {J. Chem. Theory Comput.}\
  }\textbf {\bibinfo {volume} {7}},\ \bibinfo {pages} {3412} (\bibinfo {year}
  {2011})}\BibitemShut {NoStop}%
\bibitem [{\citenamefont {Sarich}\ \emph {et~al.}(2010)\citenamefont {Sarich},
  \citenamefont {No\'{e}},\ and\ \citenamefont
  {Sch\"{u}tte}}]{SarichNoeSchuette2010}%
  \BibitemOpen
  \bibfield  {author} {\bibinfo {author} {\bibfnamefont {M.}~\bibnamefont
  {Sarich}}, \bibinfo {author} {\bibfnamefont {F.}~\bibnamefont {No\'{e}}}, \
  and\ \bibinfo {author} {\bibfnamefont {C.}~\bibnamefont {Sch\"{u}tte}},\
  }\href@noop {} {\bibfield  {journal} {\bibinfo  {journal} {Multiscale Model.
  Simul.}\ }\textbf {\bibinfo {volume} {8}},\ \bibinfo {pages} {1154} (\bibinfo
  {year} {2010})}\BibitemShut {NoStop}%
\bibitem [{\citenamefont {{W. E}}\ and\ \citenamefont {{E.
  Vanden-Eijnden}}(2006)}]{vanden2006}%
  \BibitemOpen
  \bibfield  {author} {\bibinfo {author} {\bibnamefont {{W. E}}}\ and\ \bibinfo
  {author} {\bibnamefont {{E. Vanden-Eijnden}}},\ }\href@noop {} {\bibfield
  {journal} {\bibinfo  {journal} {J. Stat. Phys.}\ }\textbf {\bibinfo {volume}
  {123}},\ \bibinfo {pages} {503} (\bibinfo {year} {2006})}\BibitemShut
  {NoStop}%
\bibitem [{\citenamefont {Metzner}\ \emph {et~al.}(2009)\citenamefont
  {Metzner}, \citenamefont {Sch\"{u}tte},\ and\ \citenamefont
  {Vanden-Eijnden}}]{metzner2009}%
  \BibitemOpen
  \bibfield  {author} {\bibinfo {author} {\bibfnamefont {P.}~\bibnamefont
  {Metzner}}, \bibinfo {author} {\bibfnamefont {C.}~\bibnamefont
  {Sch\"{u}tte}}, \ and\ \bibinfo {author} {\bibfnamefont {E.}~\bibnamefont
  {Vanden-Eijnden}},\ }\href@noop {} {\bibfield  {journal} {\bibinfo  {journal}
  {Multiscale Model. Simul.}\ }\textbf {\bibinfo {volume} {7}},\ \bibinfo
  {pages} {1192} (\bibinfo {year} {2009})}\BibitemShut {NoStop}%
\bibitem [{\citenamefont {No{\'e}}\ \emph {et~al.}(2009)\citenamefont
  {No{\'e}}, \citenamefont {Sch{\"u}tte}, \citenamefont {Vanden-Eijnden},
  \citenamefont {Reich},\ and\ \citenamefont {Weikl}}]{noe2009}%
  \BibitemOpen
  \bibfield  {author} {\bibinfo {author} {\bibfnamefont {F.}~\bibnamefont
  {No{\'e}}}, \bibinfo {author} {\bibfnamefont {C.}~\bibnamefont
  {Sch{\"u}tte}}, \bibinfo {author} {\bibfnamefont {E.}~\bibnamefont
  {Vanden-Eijnden}}, \bibinfo {author} {\bibfnamefont {L.}~\bibnamefont
  {Reich}}, \ and\ \bibinfo {author} {\bibfnamefont {T.~R.}\ \bibnamefont
  {Weikl}},\ }\href@noop {} {\bibfield  {journal} {\bibinfo  {journal} {Proc.
  Nat. Acad. Sci. USA}\ }\textbf {\bibinfo {volume} {106}},\ \bibinfo {pages}
  {19011} (\bibinfo {year} {2009})}\BibitemShut {NoStop}%
\bibitem [{\citenamefont {Voelz}\ \emph {et~al.}(2010)\citenamefont {Voelz},
  \citenamefont {Bowman}, \citenamefont {Beauchamp},\ and\ \citenamefont
  {Pande}}]{VoelzPande_JACS10_NTL9}%
  \BibitemOpen
  \bibfield  {author} {\bibinfo {author} {\bibfnamefont {V.~A.}\ \bibnamefont
  {Voelz}}, \bibinfo {author} {\bibfnamefont {G.~R.}\ \bibnamefont {Bowman}},
  \bibinfo {author} {\bibfnamefont {K.~A.}\ \bibnamefont {Beauchamp}}, \ and\
  \bibinfo {author} {\bibfnamefont {V.~S.}\ \bibnamefont {Pande}},\ }\href@noop
  {} {\bibfield  {journal} {\bibinfo  {journal} {J. Am. Chem. Soc.}\ }\textbf
  {\bibinfo {volume} {132}},\ \bibinfo {pages} {1526} (\bibinfo {year}
  {2010})}\BibitemShut {NoStop}%
\bibitem [{\citenamefont {Held}\ \emph {et~al.}(2011)\citenamefont {Held},
  \citenamefont {Metzner}, \citenamefont {Prinz},\ and\ \citenamefont
  {No{\'e}}}]{held2011}%
  \BibitemOpen
  \bibfield  {author} {\bibinfo {author} {\bibfnamefont {M.}~\bibnamefont
  {Held}}, \bibinfo {author} {\bibfnamefont {P.}~\bibnamefont {Metzner}},
  \bibinfo {author} {\bibfnamefont {J.-H.}\ \bibnamefont {Prinz}}, \ and\
  \bibinfo {author} {\bibfnamefont {F.}~\bibnamefont {No{\'e}}},\ }\href@noop
  {} {\bibfield  {journal} {\bibinfo  {journal} {Biophys. J.}\ }\textbf
  {\bibinfo {volume} {100}},\ \bibinfo {pages} {701} (\bibinfo {year}
  {2011})}\BibitemShut {NoStop}%
\bibitem [{\citenamefont {Gu}\ \emph {et~al.}(2014)\citenamefont {Gu},
  \citenamefont {Silva}, \citenamefont {Meng}, \citenamefont {Yue},\ and\
  \citenamefont {Huang}}]{huang2014}%
  \BibitemOpen
  \bibfield  {author} {\bibinfo {author} {\bibfnamefont {S.}~\bibnamefont
  {Gu}}, \bibinfo {author} {\bibfnamefont {D.-A.}\ \bibnamefont {Silva}},
  \bibinfo {author} {\bibfnamefont {L.}~\bibnamefont {Meng}}, \bibinfo {author}
  {\bibfnamefont {A.}~\bibnamefont {Yue}}, \ and\ \bibinfo {author}
  {\bibfnamefont {X.}~\bibnamefont {Huang}},\ }\href@noop {} {\bibfield
  {journal} {\bibinfo  {journal} {PLoS Comput. Biol.}\ }\textbf {\bibinfo
  {volume} {10}},\ \bibinfo {pages} {e1003767} (\bibinfo {year}
  {2014})}\BibitemShut {NoStop}%
\bibitem [{\citenamefont {Buch}\ \emph {et~al.}(2011)\citenamefont {Buch},
  \citenamefont {Giorgino},\ and\ \citenamefont {{de
  Fabritiis}}}]{BuchFabritiis_PNAS11_Binding}%
  \BibitemOpen
  \bibfield  {author} {\bibinfo {author} {\bibfnamefont {I.}~\bibnamefont
  {Buch}}, \bibinfo {author} {\bibfnamefont {T.}~\bibnamefont {Giorgino}}, \
  and\ \bibinfo {author} {\bibfnamefont {G.}~\bibnamefont {{de Fabritiis}}},\
  }\href@noop {} {\bibfield  {journal} {\bibinfo  {journal} {Proc. Natl. Acad.
  Sci. USA}\ }\textbf {\bibinfo {volume} {108}},\ \bibinfo {pages} {10184}
  (\bibinfo {year} {2011})}\BibitemShut {NoStop}%
\bibitem [{\citenamefont {Shaw}\ \emph {et~al.}(2010)\citenamefont {Shaw},
  \citenamefont {Maragakis}, \citenamefont {Lindorff-Larsen}, \citenamefont
  {Piana}, \citenamefont {Dror}, \citenamefont {Eastwood}, \citenamefont
  {Bank}, \citenamefont {Jumper}, \citenamefont {Salmon}, \citenamefont {Shan}
  \emph {et~al.}}]{shaw2010}%
  \BibitemOpen
  \bibfield  {author} {\bibinfo {author} {\bibfnamefont {D.~E.}\ \bibnamefont
  {Shaw}}, \bibinfo {author} {\bibfnamefont {P.}~\bibnamefont {Maragakis}},
  \bibinfo {author} {\bibfnamefont {K.}~\bibnamefont {Lindorff-Larsen}},
  \bibinfo {author} {\bibfnamefont {S.}~\bibnamefont {Piana}}, \bibinfo
  {author} {\bibfnamefont {R.~O.}\ \bibnamefont {Dror}}, \bibinfo {author}
  {\bibfnamefont {M.~P.}\ \bibnamefont {Eastwood}}, \bibinfo {author}
  {\bibfnamefont {J.~A.}\ \bibnamefont {Bank}}, \bibinfo {author}
  {\bibfnamefont {J.~M.}\ \bibnamefont {Jumper}}, \bibinfo {author}
  {\bibfnamefont {J.~K.}\ \bibnamefont {Salmon}}, \bibinfo {author}
  {\bibfnamefont {Y.}~\bibnamefont {Shan}},  \emph {et~al.},\ }\href@noop {}
  {\bibfield  {journal} {\bibinfo  {journal} {Science}\ }\textbf {\bibinfo
  {volume} {330}},\ \bibinfo {pages} {341} (\bibinfo {year}
  {2010})}\BibitemShut {NoStop}%
\bibitem [{\citenamefont {Torrie}\ and\ \citenamefont
  {Valleau}(1977)}]{torrie1977}%
  \BibitemOpen
  \bibfield  {author} {\bibinfo {author} {\bibfnamefont {G.}~\bibnamefont
  {Torrie}}\ and\ \bibinfo {author} {\bibfnamefont {J.}~\bibnamefont
  {Valleau}},\ }\href@noop {} {\bibfield  {journal} {\bibinfo  {journal} {J.
  Comp. Phys.}\ }\textbf {\bibinfo {volume} {23}},\ \bibinfo {pages} {187}
  (\bibinfo {year} {1977})}\BibitemShut {NoStop}%
\bibitem [{\citenamefont {Grubm{\"u}ller}(1995)}]{grubmueller1995}%
  \BibitemOpen
  \bibfield  {author} {\bibinfo {author} {\bibfnamefont {H.}~\bibnamefont
  {Grubm{\"u}ller}},\ }\href@noop {} {\bibfield  {journal} {\bibinfo  {journal}
  {Phys. Rev. E}\ }\textbf {\bibinfo {volume} {52}},\ \bibinfo {pages} {2893}
  (\bibinfo {year} {1995})}\BibitemShut {NoStop}%
\bibitem [{\citenamefont {Sugita}\ and\ \citenamefont
  {Okamoto}(1999)}]{sugita1999}%
  \BibitemOpen
  \bibfield  {author} {\bibinfo {author} {\bibfnamefont {Y.}~\bibnamefont
  {Sugita}}\ and\ \bibinfo {author} {\bibfnamefont {Y.}~\bibnamefont
  {Okamoto}},\ }\href@noop {} {\bibfield  {journal} {\bibinfo  {journal} {Chem.
  Phys. Lett.}\ }\textbf {\bibinfo {volume} {314}},\ \bibinfo {pages} {141}
  (\bibinfo {year} {1999})}\BibitemShut {NoStop}%
\bibitem [{\citenamefont {Laio}\ and\ \citenamefont
  {Parrinello}(2002)}]{laio2002}%
  \BibitemOpen
  \bibfield  {author} {\bibinfo {author} {\bibfnamefont {A.}~\bibnamefont
  {Laio}}\ and\ \bibinfo {author} {\bibfnamefont {M.}~\bibnamefont
  {Parrinello}},\ }\href@noop {} {\bibfield  {journal} {\bibinfo  {journal}
  {Proc. Natl. Acad. Sci. USA}\ }\textbf {\bibinfo {volume} {99}},\ \bibinfo
  {pages} {12562} (\bibinfo {year} {2002})}\BibitemShut {NoStop}%
\bibitem [{\citenamefont {Eyring}(1935)}]{eyring1935}%
  \BibitemOpen
  \bibfield  {author} {\bibinfo {author} {\bibfnamefont {H.}~\bibnamefont
  {Eyring}},\ }\href@noop {} {\bibfield  {journal} {\bibinfo  {journal} {J.
  Chem. Phys.}\ }\textbf {\bibinfo {volume} {3}},\ \bibinfo {pages} {107}
  (\bibinfo {year} {1935})}\BibitemShut {NoStop}%
\bibitem [{\citenamefont {Kramers}(1940)}]{kramers1940}%
  \BibitemOpen
  \bibfield  {author} {\bibinfo {author} {\bibfnamefont {H.}~\bibnamefont
  {Kramers}},\ }\href@noop {} {\bibfield  {journal} {\bibinfo  {journal}
  {Physica}\ }\textbf {\bibinfo {volume} {7}},\ \bibinfo {pages} {284 }
  (\bibinfo {year} {1940})}\BibitemShut {NoStop}%
\bibitem [{\citenamefont {Best}\ and\ \citenamefont
  {Hummer}(2010)}]{BestHummer_PNAS09_Diffusion}%
  \BibitemOpen
  \bibfield  {author} {\bibinfo {author} {\bibfnamefont {R.~B.}\ \bibnamefont
  {Best}}\ and\ \bibinfo {author} {\bibfnamefont {G.}~\bibnamefont {Hummer}},\
  }\href@noop {} {\bibfield  {journal} {\bibinfo  {journal} {Proc. Natl. Acad.
  Sci. USA}\ }\textbf {\bibinfo {volume} {107}},\ \bibinfo {pages} {1088}
  (\bibinfo {year} {2010})}\BibitemShut {NoStop}%
\bibitem [{\citenamefont {Tiwary}\ and\ \citenamefont
  {Parrinello}(2013)}]{tiwary2013}%
  \BibitemOpen
  \bibfield  {author} {\bibinfo {author} {\bibfnamefont {P.}~\bibnamefont
  {Tiwary}}\ and\ \bibinfo {author} {\bibfnamefont {M.}~\bibnamefont
  {Parrinello}},\ }\href@noop {} {\bibfield  {journal} {\bibinfo  {journal}
  {Phys. Rev. Lett.}\ }\textbf {\bibinfo {volume} {111}},\ \bibinfo {pages}
  {230602} (\bibinfo {year} {2013})}\BibitemShut {NoStop}%
\bibitem [{\citenamefont {Rosta}\ and\ \citenamefont
  {Hummer}(2015)}]{rosta2015}%
  \BibitemOpen
  \bibfield  {author} {\bibinfo {author} {\bibfnamefont {E.}~\bibnamefont
  {Rosta}}\ and\ \bibinfo {author} {\bibfnamefont {G.}~\bibnamefont {Hummer}},\
  }\href@noop {} {\bibfield  {journal} {\bibinfo  {journal} {Journal of
  Chemical Theory and Computation}\ }\textbf {\bibinfo {volume} {11}},\
  \bibinfo {pages} {276} (\bibinfo {year} {2015})}\BibitemShut {NoStop}%
\bibitem [{\citenamefont {Wu}\ and\ \citenamefont {No\'{e}}(2014)}]{WuNoe2014}%
  \BibitemOpen
  \bibfield  {author} {\bibinfo {author} {\bibfnamefont {H.}~\bibnamefont
  {Wu}}\ and\ \bibinfo {author} {\bibfnamefont {F.}~\bibnamefont {No\'{e}}},\
  }\href@noop {} {\bibfield  {journal} {\bibinfo  {journal} {Multiscale
  Modeling \& Simulation}\ }\textbf {\bibinfo {volume} {12}},\ \bibinfo {pages}
  {25} (\bibinfo {year} {2014})}\BibitemShut {NoStop}%
\bibitem [{\citenamefont {Wu}\ \emph {et~al.}(2014)\citenamefont {Wu},
  \citenamefont {Mey}, \citenamefont {Rosta},\ and\ \citenamefont
  {No\'{e}}}]{WuMeyRostaNoe2014}%
  \BibitemOpen
  \bibfield  {author} {\bibinfo {author} {\bibfnamefont {H.}~\bibnamefont
  {Wu}}, \bibinfo {author} {\bibfnamefont {A.}~\bibnamefont {Mey}}, \bibinfo
  {author} {\bibfnamefont {E.}~\bibnamefont {Rosta}}, \ and\ \bibinfo {author}
  {\bibfnamefont {F.}~\bibnamefont {No\'{e}}},\ }\href@noop {} {\bibfield
  {journal} {\bibinfo  {journal} {The Journal of Chemical Physics}\ }\textbf
  {\bibinfo {volume} {141}},\ \bibinfo {eid} {214106} (\bibinfo {year}
  {2014})}\BibitemShut {NoStop}%
\bibitem [{\citenamefont {Mey}\ \emph {et~al.}(2014)\citenamefont {Mey},
  \citenamefont {Wu},\ and\ \citenamefont {No\'e}}]{MeyWuNoe2014}%
  \BibitemOpen
  \bibfield  {author} {\bibinfo {author} {\bibfnamefont {A.}~\bibnamefont
  {Mey}}, \bibinfo {author} {\bibfnamefont {H.}~\bibnamefont {Wu}}, \ and\
  \bibinfo {author} {\bibfnamefont {F.}~\bibnamefont {No\'e}},\ }\href
  {\doibase 10.1103/PhysRevX.4.041018} {\bibfield  {journal} {\bibinfo
  {journal} {Phys. Rev. X}\ }\textbf {\bibinfo {volume} {4}},\ \bibinfo {pages}
  {041018} (\bibinfo {year} {2014})}\BibitemShut {NoStop}%
\bibitem [{\citenamefont {Bolhuis}\ \emph {et~al.}(2002)\citenamefont
  {Bolhuis}, \citenamefont {Chandler}, \citenamefont {Dellago},\ and\
  \citenamefont {Geissler}}]{bolhuis2002}%
  \BibitemOpen
  \bibfield  {author} {\bibinfo {author} {\bibfnamefont {P.~G.}\ \bibnamefont
  {Bolhuis}}, \bibinfo {author} {\bibfnamefont {D.}~\bibnamefont {Chandler}},
  \bibinfo {author} {\bibfnamefont {C.}~\bibnamefont {Dellago}}, \ and\
  \bibinfo {author} {\bibfnamefont {P.~L.}\ \bibnamefont {Geissler}},\
  }\href@noop {} {\bibfield  {journal} {\bibinfo  {journal} {Annu. Rev. Phys.
  Chem.}\ }\textbf {\bibinfo {volume} {53}},\ \bibinfo {pages} {291} (\bibinfo
  {year} {2002})}\BibitemShut {NoStop}%
\bibitem [{\citenamefont {Faradjian}\ and\ \citenamefont
  {Elber}(2004)}]{faradjian2004}%
  \BibitemOpen
  \bibfield  {author} {\bibinfo {author} {\bibfnamefont {A.~K.}\ \bibnamefont
  {Faradjian}}\ and\ \bibinfo {author} {\bibfnamefont {R.}~\bibnamefont
  {Elber}},\ }\href@noop {} {\bibfield  {journal} {\bibinfo  {journal} {J.
  Chem. Phys.}\ }\textbf {\bibinfo {volume} {120}},\ \bibinfo {pages} {10880}
  (\bibinfo {year} {2004})}\BibitemShut {NoStop}%
\bibitem [{\citenamefont {van Erp}\ \emph {et~al.}(2003)\citenamefont {van
  Erp}, \citenamefont {Moroni},\ and\ \citenamefont {Bolhuis}}]{vanErp2003}%
  \BibitemOpen
  \bibfield  {author} {\bibinfo {author} {\bibfnamefont {T.~S.}\ \bibnamefont
  {van Erp}}, \bibinfo {author} {\bibfnamefont {D.}~\bibnamefont {Moroni}}, \
  and\ \bibinfo {author} {\bibfnamefont {P.~G.}\ \bibnamefont {Bolhuis}},\
  }\href@noop {} {\bibfield  {journal} {\bibinfo  {journal} {J. Chem. Phys.}\
  }\textbf {\bibinfo {volume} {118}} (\bibinfo {year} {2003})}\BibitemShut
  {NoStop}%
\bibitem [{\citenamefont {Du}\ and\ \citenamefont {Bolhuis}(2013)}]{du2013}%
  \BibitemOpen
  \bibfield  {author} {\bibinfo {author} {\bibfnamefont {W.-N.}\ \bibnamefont
  {Du}}\ and\ \bibinfo {author} {\bibfnamefont {P.~G.}\ \bibnamefont
  {Bolhuis}},\ }\href@noop {} {\bibfield  {journal} {\bibinfo  {journal} {J.
  Chem. Phys.}\ }\textbf {\bibinfo {volume} {139}},\ \bibinfo {eid} {044105}
  (\bibinfo {year} {2013})}\BibitemShut {NoStop}%
\bibitem [{\citenamefont {Van~Gunsteren}\ and\ \citenamefont
  {Berendsen}(1988)}]{vanGunsteren1988}%
  \BibitemOpen
  \bibfield  {author} {\bibinfo {author} {\bibfnamefont {W.~F.}\ \bibnamefont
  {Van~Gunsteren}}\ and\ \bibinfo {author} {\bibfnamefont {H.~J.~C.}\
  \bibnamefont {Berendsen}},\ }\href@noop {} {\bibfield  {journal} {\bibinfo
  {journal} {Mol. Simulat.}\ }\textbf {\bibinfo {volume} {1}},\ \bibinfo
  {pages} {173} (\bibinfo {year} {1988})}\BibitemShut {NoStop}%
\bibitem [{\citenamefont {Tuckerman}\ \emph {et~al.}(1992)\citenamefont
  {Tuckerman}, \citenamefont {Berne},\ and\ \citenamefont
  {Martyna}}]{tuckerman1992}%
  \BibitemOpen
  \bibfield  {author} {\bibinfo {author} {\bibfnamefont {M.}~\bibnamefont
  {Tuckerman}}, \bibinfo {author} {\bibfnamefont {B.~J.}\ \bibnamefont
  {Berne}}, \ and\ \bibinfo {author} {\bibfnamefont {G.~J.}\ \bibnamefont
  {Martyna}},\ }\href@noop {} {\bibfield  {journal} {\bibinfo  {journal} {J.
  Chem. Phys.}\ }\textbf {\bibinfo {volume} {97}},\ \bibinfo {pages} {1990}
  (\bibinfo {year} {1992})}\BibitemShut {NoStop}%
\bibitem [{\citenamefont {Sriraman}\ \emph {et~al.}(2005)\citenamefont
  {Sriraman}, \citenamefont {Kevrekidis},\ and\ \citenamefont
  {Hummer}}]{SriramanKevrekidisHummer_JPCB109_6479}%
  \BibitemOpen
  \bibfield  {author} {\bibinfo {author} {\bibfnamefont {S.}~\bibnamefont
  {Sriraman}}, \bibinfo {author} {\bibfnamefont {I.~G.}\ \bibnamefont
  {Kevrekidis}}, \ and\ \bibinfo {author} {\bibfnamefont {G.}~\bibnamefont
  {Hummer}},\ }\href@noop {} {\bibfield  {journal} {\bibinfo  {journal} {J.
  Phys. Chem. B}\ }\textbf {\bibinfo {volume} {109}},\ \bibinfo {pages} {6479}
  (\bibinfo {year} {2005})}\BibitemShut {NoStop}%
\bibitem [{\citenamefont {No{\'e}}(2008)}]{noe2008}%
  \BibitemOpen
  \bibfield  {author} {\bibinfo {author} {\bibfnamefont {F.}~\bibnamefont
  {No{\'e}}},\ }\href@noop {} {\bibfield  {journal} {\bibinfo  {journal} {J.
  Chem. Phys.}\ }\textbf {\bibinfo {volume} {128}},\ \bibinfo {pages} {244103}
  (\bibinfo {year} {2008})}\BibitemShut {NoStop}%
\bibitem [{\citenamefont {Wang}\ \emph {et~al.}(2013)\citenamefont {Wang},
  \citenamefont {Chodera}, \citenamefont {Yang},\ and\ \citenamefont
  {Shirts}}]{wang2013}%
  \BibitemOpen
  \bibfield  {author} {\bibinfo {author} {\bibfnamefont {K.}~\bibnamefont
  {Wang}}, \bibinfo {author} {\bibfnamefont {J.~D.}\ \bibnamefont {Chodera}},
  \bibinfo {author} {\bibfnamefont {Y.}~\bibnamefont {Yang}}, \ and\ \bibinfo
  {author} {\bibfnamefont {M.~R.}\ \bibnamefont {Shirts}},\ }\href@noop {}
  {\bibfield  {journal} {\bibinfo  {journal} {J. Comput. Aided Mol. Des.}\
  }\textbf {\bibinfo {volume} {27}},\ \bibinfo {pages} {989} (\bibinfo {year}
  {2013})}\BibitemShut {NoStop}%
\bibitem [{\citenamefont {Souaille}\ and\ \citenamefont
  {Roux}(2001)}]{SouailleRoux_CPC01_WHAM}%
  \BibitemOpen
  \bibfield  {author} {\bibinfo {author} {\bibfnamefont {M.}~\bibnamefont
  {Souaille}}\ and\ \bibinfo {author} {\bibfnamefont {B.}~\bibnamefont
  {Roux}},\ }\href@noop {} {\bibfield  {journal} {\bibinfo  {journal} {Comput.
  Phys. Commun.}\ }\textbf {\bibinfo {volume} {135}},\ \bibinfo {pages} {40}
  (\bibinfo {year} {2001})}\BibitemShut {NoStop}%
\bibitem [{\citenamefont {Zwanzig}(1973)}]{zwanzig1973}%
  \BibitemOpen
  \bibfield  {author} {\bibinfo {author} {\bibfnamefont {R.}~\bibnamefont
  {Zwanzig}},\ }\href@noop {} {\bibfield  {journal} {\bibinfo  {journal} {J.
  Stat. Phys.}\ }\textbf {\bibinfo {volume} {9}},\ \bibinfo {pages} {215}
  (\bibinfo {year} {1973})}\BibitemShut {NoStop}%
\bibitem [{\citenamefont {Gardiner}(1986)}]{gardiner1986}%
  \BibitemOpen
  \bibfield  {author} {\bibinfo {author} {\bibfnamefont {C.~W.}\ \bibnamefont
  {Gardiner}},\ }\href@noop {} {\bibfield  {journal} {\bibinfo  {journal}
  {Applied Optics}\ }\textbf {\bibinfo {volume} {25}},\ \bibinfo {pages} {3145}
  (\bibinfo {year} {1986})}\BibitemShut {NoStop}%
\bibitem [{\citenamefont {Sch{\"u}tte}(1999)}]{schuette1999}%
  \BibitemOpen
  \bibfield  {author} {\bibinfo {author} {\bibfnamefont {C.}~\bibnamefont
  {Sch{\"u}tte}},\ }\href@noop {} {\enquote {\bibinfo {title} {Conformational
  dynamics: Modeling, theory, algorithm, and application to biomolecules},}\ }
  (\bibinfo {year} {1999}),\ \bibinfo {note} {habilitation thesis}\BibitemShut
  {NoStop}%
\bibitem [{\citenamefont {Sch{\"u}tte}\ and\ \citenamefont
  {Huisinga}(2000)}]{schuette2000}%
  \BibitemOpen
  \bibfield  {author} {\bibinfo {author} {\bibfnamefont {C.}~\bibnamefont
  {Sch{\"u}tte}}\ and\ \bibinfo {author} {\bibfnamefont {W.}~\bibnamefont
  {Huisinga}},\ }\enquote {\bibinfo {title} {On conformational dynamics induced
  by langevin processes},}\ in\ \href@noop {} {\emph {\bibinfo {booktitle}
  {Equadiff 99}}}\ (\bibinfo {year} {2000})\ Chap.\ \bibinfo {chapter} {234},
  pp.\ \bibinfo {pages} {1247--1262}\BibitemShut {NoStop}%
\bibitem [{\citenamefont {Wang}\ and\ \citenamefont {Landau}(2001)}]{wang2001}%
  \BibitemOpen
  \bibfield  {author} {\bibinfo {author} {\bibfnamefont {F.}~\bibnamefont
  {Wang}}\ and\ \bibinfo {author} {\bibfnamefont {D.~P.}\ \bibnamefont
  {Landau}},\ }\href@noop {} {\bibfield  {journal} {\bibinfo  {journal} {Phys.
  Rev. Lett.}\ }\textbf {\bibinfo {volume} {86}},\ \bibinfo {pages} {2050}
  (\bibinfo {year} {2001})}\BibitemShut {NoStop}%
\bibitem [{\citenamefont {Trebst}\ and\ \citenamefont
  {Troyer}(2006)}]{trebst2006}%
  \BibitemOpen
  \bibfield  {author} {\bibinfo {author} {\bibfnamefont {S.}~\bibnamefont
  {Trebst}}\ and\ \bibinfo {author} {\bibfnamefont {M.}~\bibnamefont
  {Troyer}},\ }in\ \href@noop {} {\emph {\bibinfo {booktitle} {Computer
  Simulations in Condensed Matter: From Materials to Chemical Biology. Volume
  1}}},\ \bibinfo {editor} {edited by\ \bibinfo {editor} {\bibfnamefont
  {M.}~\bibnamefont {Ferrario}}, \bibinfo {editor} {\bibfnamefont
  {G.}~\bibnamefont {Ciccotti}}, \ and\ \bibinfo {editor} {\bibfnamefont
  {K.}~\bibnamefont {Binder}}}\ (\bibinfo  {publisher} {Springer},\ \bibinfo
  {year} {2006})\BibitemShut {NoStop}%
\bibitem [{\citenamefont {Bennett}(1976)}]{bennett1976}%
  \BibitemOpen
  \bibfield  {author} {\bibinfo {author} {\bibfnamefont {C.~H.}\ \bibnamefont
  {Bennett}},\ }\href@noop {} {\ \textbf {\bibinfo {volume} {22}},\ \bibinfo
  {pages} {245 } (\bibinfo {year} {1976})}\BibitemShut {NoStop}%
\bibitem [{\citenamefont {Ferrenberg}\ and\ \citenamefont
  {Swendsen}(1989)}]{ferrenberg1989}%
  \BibitemOpen
  \bibfield  {author} {\bibinfo {author} {\bibfnamefont {A.~M.}\ \bibnamefont
  {Ferrenberg}}\ and\ \bibinfo {author} {\bibfnamefont {R.~H.}\ \bibnamefont
  {Swendsen}},\ }\href@noop {} {\bibfield  {journal} {\bibinfo  {journal}
  {Physical Review Letters}\ }\textbf {\bibinfo {volume} {63}},\ \bibinfo
  {pages} {1195} (\bibinfo {year} {1989})}\BibitemShut {NoStop}%
\bibitem [{\citenamefont {Kumar}\ \emph {et~al.}(1995)\citenamefont {Kumar},
  \citenamefont {Rosenberg}, \citenamefont {Bouzida}, \citenamefont
  {Swendsen},\ and\ \citenamefont {Kollman}}]{kumar1995}%
  \BibitemOpen
  \bibfield  {author} {\bibinfo {author} {\bibfnamefont {S.}~\bibnamefont
  {Kumar}}, \bibinfo {author} {\bibfnamefont {J.}~\bibnamefont {Rosenberg}},
  \bibinfo {author} {\bibfnamefont {D.}~\bibnamefont {Bouzida}}, \bibinfo
  {author} {\bibfnamefont {R.}~\bibnamefont {Swendsen}}, \ and\ \bibinfo
  {author} {\bibfnamefont {P.}~\bibnamefont {Kollman}},\ }\href@noop {}
  {\bibfield  {journal} {\bibinfo  {journal} {J. Comp. Chem.}\ }\textbf
  {\bibinfo {volume} {16}},\ \bibinfo {pages} {1339} (\bibinfo {year}
  {1995})}\BibitemShut {NoStop}%
\bibitem [{\citenamefont {Tan}(2004)}]{tan2004}%
  \BibitemOpen
  \bibfield  {author} {\bibinfo {author} {\bibfnamefont {Z.}~\bibnamefont
  {Tan}},\ }\href@noop {} {\bibfield  {journal} {\bibinfo  {journal} {J. Amer.
  Statist. Assoc.}\ }\textbf {\bibinfo {volume} {99}},\ \bibinfo {pages} {1027}
  (\bibinfo {year} {2004})}\BibitemShut {NoStop}%
\bibitem [{\citenamefont {Shirts}\ and\ \citenamefont
  {Chodera}(2008)}]{shirts2008}%
  \BibitemOpen
  \bibfield  {author} {\bibinfo {author} {\bibfnamefont {M.~R.}\ \bibnamefont
  {Shirts}}\ and\ \bibinfo {author} {\bibfnamefont {J.~D.}\ \bibnamefont
  {Chodera}},\ }\href@noop {} {\bibfield  {journal} {\bibinfo  {journal} {J.
  Chem. Phys.}\ }\textbf {\bibinfo {volume} {129}},\ \bibinfo {eid} {124105}
  (\bibinfo {year} {2008})}\BibitemShut {NoStop}%
\bibitem [{\citenamefont {Trendelkamp-Schroer}\ \emph
  {et~al.}(tted)\citenamefont {Trendelkamp-Schroer}, \citenamefont {Wu},
  \citenamefont {Paul},\ and\ \citenamefont {No\'{e}}}]{TrendelkampWuNoe2015}%
  \BibitemOpen
  \bibfield  {author} {\bibinfo {author} {\bibfnamefont {B.}~\bibnamefont
  {Trendelkamp-Schroer}}, \bibinfo {author} {\bibfnamefont {H.}~\bibnamefont
  {Wu}}, \bibinfo {author} {\bibfnamefont {F.}~\bibnamefont {Paul}}, \ and\
  \bibinfo {author} {\bibfnamefont {F.}~\bibnamefont {No\'{e}}},\ }\href@noop
  {} {\bibfield  {journal} {\bibinfo  {journal} {J. Chem. Phys.}\ } (\bibinfo
  {year} {2015, submitted})}\BibitemShut {NoStop}%
\bibitem [{\citenamefont {Kumar}\ \emph {et~al.}(1992)\citenamefont {Kumar},
  \citenamefont {Rosenberg}, \citenamefont {Bouzida}, \citenamefont
  {Swendsen},\ and\ \citenamefont {Kollman}}]{kumar1992}%
  \BibitemOpen
  \bibfield  {author} {\bibinfo {author} {\bibfnamefont {S.}~\bibnamefont
  {Kumar}}, \bibinfo {author} {\bibfnamefont {J.~M.}\ \bibnamefont
  {Rosenberg}}, \bibinfo {author} {\bibfnamefont {D.}~\bibnamefont {Bouzida}},
  \bibinfo {author} {\bibfnamefont {R.~H.}\ \bibnamefont {Swendsen}}, \ and\
  \bibinfo {author} {\bibfnamefont {P.~A.}\ \bibnamefont {Kollman}},\
  }\href@noop {} {\bibfield  {journal} {\bibinfo  {journal} {J. Comput. Chem.}\
  }\textbf {\bibinfo {volume} {13}},\ \bibinfo {pages} {1011} (\bibinfo {year}
  {1992})}\BibitemShut {NoStop}%
\bibitem [{\citenamefont {Efron}\ and\ \citenamefont
  {Tibshirani}(1994)}]{efron1994}%
  \BibitemOpen
  \bibfield  {author} {\bibinfo {author} {\bibfnamefont {B.}~\bibnamefont
  {Efron}}\ and\ \bibinfo {author} {\bibfnamefont {R.~J.}\ \bibnamefont
  {Tibshirani}},\ }\href@noop {} {\emph {\bibinfo {title} {An introduction to
  the bootstrap}}}\ (\bibinfo  {publisher} {CRC press},\ \bibinfo {year}
  {1994})\BibitemShut {NoStop}%
\bibitem [{\citenamefont {Montgomery~Pettitt}\ and\ \citenamefont
  {Karplus}(1985)}]{montgomery1985}%
  \BibitemOpen
  \bibfield  {author} {\bibinfo {author} {\bibfnamefont {B.}~\bibnamefont
  {Montgomery~Pettitt}}\ and\ \bibinfo {author} {\bibfnamefont
  {M.}~\bibnamefont {Karplus}},\ }\href@noop {} {\bibfield  {journal} {\bibinfo
   {journal} {Chem. Phys. Lett.}\ }\textbf {\bibinfo {volume} {121}},\ \bibinfo
  {pages} {194} (\bibinfo {year} {1985})}\BibitemShut {NoStop}%
\bibitem [{\citenamefont {Anderson}\ and\ \citenamefont
  {Hermans}(1988)}]{anderson1988}%
  \BibitemOpen
  \bibfield  {author} {\bibinfo {author} {\bibfnamefont {A.~G.}\ \bibnamefont
  {Anderson}}\ and\ \bibinfo {author} {\bibfnamefont {J.}~\bibnamefont
  {Hermans}},\ }\href@noop {} {\bibfield  {journal} {\bibinfo  {journal}
  {Proteins}\ }\textbf {\bibinfo {volume} {3}},\ \bibinfo {pages} {262}
  (\bibinfo {year} {1988})}\BibitemShut {NoStop}%
\bibitem [{\citenamefont {Tobias}\ and\ \citenamefont
  {Brooks~III}(1992)}]{tobias1992}%
  \BibitemOpen
  \bibfield  {author} {\bibinfo {author} {\bibfnamefont {D.~J.}\ \bibnamefont
  {Tobias}}\ and\ \bibinfo {author} {\bibfnamefont {C.~L.}\ \bibnamefont
  {Brooks~III}},\ }\href@noop {} {\bibfield  {journal} {\bibinfo  {journal} {J.
  Phys. Chem.}\ }\textbf {\bibinfo {volume} {96}},\ \bibinfo {pages} {3864}
  (\bibinfo {year} {1992})}\BibitemShut {NoStop}%
\bibitem [{\citenamefont {Chodera}\ \emph {et~al.}(2006)\citenamefont
  {Chodera}, \citenamefont {Swope}, \citenamefont {Pitera},\ and\ \citenamefont
  {Dill}}]{chodera2006}%
  \BibitemOpen
  \bibfield  {author} {\bibinfo {author} {\bibfnamefont {J.~D.}\ \bibnamefont
  {Chodera}}, \bibinfo {author} {\bibfnamefont {W.~C.}\ \bibnamefont {Swope}},
  \bibinfo {author} {\bibfnamefont {J.~W.}\ \bibnamefont {Pitera}}, \ and\
  \bibinfo {author} {\bibfnamefont {K.~A.}\ \bibnamefont {Dill}},\ }\href@noop
  {} {\bibfield  {journal} {\bibinfo  {journal} {Multiscale Model. Simul.}\
  }\textbf {\bibinfo {volume} {5}},\ \bibinfo {pages} {1214} (\bibinfo {year}
  {2006})}\BibitemShut {NoStop}%
\bibitem [{\citenamefont {Du}\ \emph {et~al.}(2011)\citenamefont {Du},
  \citenamefont {Marino},\ and\ \citenamefont {Bolhuis}}]{du2011}%
  \BibitemOpen
  \bibfield  {author} {\bibinfo {author} {\bibfnamefont {W.-N.}\ \bibnamefont
  {Du}}, \bibinfo {author} {\bibfnamefont {K.~A.}\ \bibnamefont {Marino}}, \
  and\ \bibinfo {author} {\bibfnamefont {P.~G.}\ \bibnamefont {Bolhuis}},\
  }\href@noop {} {\bibfield  {journal} {\bibinfo  {journal} {J. Chem. Phys.}\
  }\textbf {\bibinfo {volume} {135}},\ \bibinfo {pages} {145102} (\bibinfo
  {year} {2011})}\BibitemShut {NoStop}%
\bibitem [{\citenamefont {Eastman}\ \emph {et~al.}(2013)\citenamefont
  {Eastman}, \citenamefont {Friedrichs}, \citenamefont {Chodera}, \citenamefont
  {Radmer}, \citenamefont {Bruns}, \citenamefont {Ku}, \citenamefont
  {Beauchamp}, \citenamefont {Lane}, \citenamefont {Wang}, \citenamefont
  {Shukla}, \citenamefont {Tye}, \citenamefont {Houston}, \citenamefont
  {Stich}, \citenamefont {Klein}, \citenamefont {Shirts},\ and\ \citenamefont
  {Pande}}]{eastman2013}%
  \BibitemOpen
  \bibfield  {author} {\bibinfo {author} {\bibfnamefont {P.}~\bibnamefont
  {Eastman}}, \bibinfo {author} {\bibfnamefont {M.~S.}\ \bibnamefont
  {Friedrichs}}, \bibinfo {author} {\bibfnamefont {J.~D.}\ \bibnamefont
  {Chodera}}, \bibinfo {author} {\bibfnamefont {R.~J.}\ \bibnamefont {Radmer}},
  \bibinfo {author} {\bibfnamefont {C.~M.}\ \bibnamefont {Bruns}}, \bibinfo
  {author} {\bibfnamefont {J.~P.}\ \bibnamefont {Ku}}, \bibinfo {author}
  {\bibfnamefont {K.~A.}\ \bibnamefont {Beauchamp}}, \bibinfo {author}
  {\bibfnamefont {T.~J.}\ \bibnamefont {Lane}}, \bibinfo {author}
  {\bibfnamefont {L.-P.}\ \bibnamefont {Wang}}, \bibinfo {author}
  {\bibfnamefont {D.}~\bibnamefont {Shukla}}, \bibinfo {author} {\bibfnamefont
  {T.}~\bibnamefont {Tye}}, \bibinfo {author} {\bibfnamefont {M.}~\bibnamefont
  {Houston}}, \bibinfo {author} {\bibfnamefont {T.}~\bibnamefont {Stich}},
  \bibinfo {author} {\bibfnamefont {C.}~\bibnamefont {Klein}}, \bibinfo
  {author} {\bibfnamefont {M.~R.}\ \bibnamefont {Shirts}}, \ and\ \bibinfo
  {author} {\bibfnamefont {V.~S.}\ \bibnamefont {Pande}},\ }\href@noop {}
  {\bibfield  {journal} {\bibinfo  {journal} {J. Chem. Theory Comput.}\
  }\textbf {\bibinfo {volume} {9}},\ \bibinfo {pages} {461} (\bibinfo {year}
  {2013})}\BibitemShut {NoStop}%
\bibitem [{\citenamefont {Lindorff-Larsen}\ \emph {et~al.}(2010)\citenamefont
  {Lindorff-Larsen}, \citenamefont {Piana}, \citenamefont {Palmo},
  \citenamefont {Maragakis}, \citenamefont {Klepeis}, \citenamefont {Dror},\
  and\ \citenamefont {Shaw}}]{lindorff2010}%
  \BibitemOpen
  \bibfield  {author} {\bibinfo {author} {\bibfnamefont {K.}~\bibnamefont
  {Lindorff-Larsen}}, \bibinfo {author} {\bibfnamefont {S.}~\bibnamefont
  {Piana}}, \bibinfo {author} {\bibfnamefont {K.}~\bibnamefont {Palmo}},
  \bibinfo {author} {\bibfnamefont {P.}~\bibnamefont {Maragakis}}, \bibinfo
  {author} {\bibfnamefont {J.~L.}\ \bibnamefont {Klepeis}}, \bibinfo {author}
  {\bibfnamefont {R.~O.}\ \bibnamefont {Dror}}, \ and\ \bibinfo {author}
  {\bibfnamefont {D.~E.}\ \bibnamefont {Shaw}},\ }\href@noop {} {\bibfield
  {journal} {\bibinfo  {journal} {Proteins}\ }\textbf {\bibinfo {volume}
  {78}},\ \bibinfo {pages} {1950} (\bibinfo {year} {2010})}\BibitemShut
  {NoStop}%
\bibitem [{\citenamefont {Jorgensen}\ \emph {et~al.}(1983)\citenamefont
  {Jorgensen}, \citenamefont {Chandrasekhar}, \citenamefont {Madura},
  \citenamefont {Impey},\ and\ \citenamefont {Klein}}]{jorgensen1983}%
  \BibitemOpen
  \bibfield  {author} {\bibinfo {author} {\bibfnamefont {W.~L.}\ \bibnamefont
  {Jorgensen}}, \bibinfo {author} {\bibfnamefont {J.}~\bibnamefont
  {Chandrasekhar}}, \bibinfo {author} {\bibfnamefont {J.~D.}\ \bibnamefont
  {Madura}}, \bibinfo {author} {\bibfnamefont {R.~W.}\ \bibnamefont {Impey}}, \
  and\ \bibinfo {author} {\bibfnamefont {M.~L.}\ \bibnamefont {Klein}},\
  }\href@noop {} {\bibfield  {journal} {\bibinfo  {journal} {J. Chem. Phys.}\
  }\textbf {\bibinfo {volume} {79}},\ \bibinfo {pages} {926} (\bibinfo {year}
  {1983})}\BibitemShut {NoStop}%
\bibitem [{\citenamefont {Tummino}\ and\ \citenamefont
  {Copeland}(2008)}]{tummino2008}%
  \BibitemOpen
  \bibfield  {author} {\bibinfo {author} {\bibfnamefont {P.~J.}\ \bibnamefont
  {Tummino}}\ and\ \bibinfo {author} {\bibfnamefont {R.~A.}\ \bibnamefont
  {Copeland}},\ }\href@noop {} {\bibfield  {journal} {\bibinfo  {journal}
  {Biochemistry}\ }\textbf {\bibinfo {volume} {47}},\ \bibinfo {pages} {5481}
  (\bibinfo {year} {2008})}\BibitemShut {NoStop}%
\bibitem [{\citenamefont {Bowman}\ \emph {et~al.}(2010)\citenamefont {Bowman},
  \citenamefont {Ensign},\ and\ \citenamefont
  {Pande}}]{BowmanEnsignPande_JCTC2010_AdaptiveSampling}%
  \BibitemOpen
  \bibfield  {author} {\bibinfo {author} {\bibfnamefont {G.~R.}\ \bibnamefont
  {Bowman}}, \bibinfo {author} {\bibfnamefont {D.~L.}\ \bibnamefont {Ensign}},
  \ and\ \bibinfo {author} {\bibfnamefont {V.~S.}\ \bibnamefont {Pande}},\
  }\href@noop {} {\bibfield  {journal} {\bibinfo  {journal} {J. Chem. Theory
  Comput.}\ }\textbf {\bibinfo {volume} {6}},\ \bibinfo {pages} {787} (\bibinfo
  {year} {2010})}\BibitemShut {NoStop}%
\bibitem [{\citenamefont {Doerr}\ and\ \citenamefont
  {De~Fabritiis}(2014)}]{doerr2014}%
  \BibitemOpen
  \bibfield  {author} {\bibinfo {author} {\bibfnamefont {S.}~\bibnamefont
  {Doerr}}\ and\ \bibinfo {author} {\bibfnamefont {G.}~\bibnamefont
  {De~Fabritiis}},\ }\href@noop {} {\bibfield  {journal} {\bibinfo  {journal}
  {J. Chem. Theory Comput.}\ }\textbf {\bibinfo {volume} {10}},\ \bibinfo
  {pages} {2064} (\bibinfo {year} {2014})}\BibitemShut {NoStop}%
\bibitem [{\citenamefont {Kloeden}\ and\ \citenamefont
  {Platen}(1992)}]{kloeden1992}%
  \BibitemOpen
  \bibfield  {author} {\bibinfo {author} {\bibfnamefont {P.~E.}\ \bibnamefont
  {Kloeden}}\ and\ \bibinfo {author} {\bibfnamefont {E.}~\bibnamefont
  {Platen}},\ }\href@noop {} {\emph {\bibinfo {title} {Numerical solution of
  stochastic differential equations}}},\ Vol.~\bibinfo {volume} {23}\ (\bibinfo
   {publisher} {Springer},\ \bibinfo {year} {1992})\BibitemShut {NoStop}%
\bibitem [{\citenamefont {Hoel}\ \emph {et~al.}(1986)\citenamefont {Hoel},
  \citenamefont {Port},\ and\ \citenamefont {Stone}}]{hoel1986}%
  \BibitemOpen
  \bibfield  {author} {\bibinfo {author} {\bibfnamefont {P.~G.}\ \bibnamefont
  {Hoel}}, \bibinfo {author} {\bibfnamefont {S.~C.}\ \bibnamefont {Port}}, \
  and\ \bibinfo {author} {\bibfnamefont {C.~J.}\ \bibnamefont {Stone}},\
  }\href@noop {} {\emph {\bibinfo {title} {Introduction to stochastic
  processes}}}\ (\bibinfo  {publisher} {Waveland Press},\ \bibinfo {year}
  {1986})\BibitemShut {NoStop}%
\end{thebibliography}
